\pdfoutput=1

\documentclass[nonacm]{acmart}

\renewcommand\footnotetextcopyrightpermission[1]{}
\usepackage{etoolbox}
\usepackage{multirow}
\usepackage{adjustbox}
\AtBeginDocument{%
  }

\usepackage{etoolbox}

\makeatletter
\patchcmd{\@mktitle}{\@titlefont\@title}{\centering\@titlefont\@title}{}{}

\preto{\@mkauthors}{\begin{center}}
\appto{\@mkauthors}{\end{center}}

\patchcmd{\@mkauthors@i}{\begin{tabular}[t]{@{}l@{}}}{\begin{tabular}[t]{@{}c@{}}}{}{}
\patchcmd{\@mkauthors@ii}{\begin{tabular}[t]{@{}l@{}}}{\begin{tabular}[t]{@{}c@{}}}{}{}
\patchcmd{\@mkauthors@iii}{\begin{tabular}[t]{@{}l@{}}}{\begin{tabular}[t]{@{}c@{}}}{}{}

\def\@authorfont{\Large\normalfont\bfseries}
\def\@affiliationfont{\normalsize\normalfont}

\let\@authorsaddresses\@empty
\makeatother

\begin{document}
\title{\centering Leveraging Language Semantics for Collaborative Filtering with TextGCN and TextGCN-MLP: Zero-Shot vs In-Domain Performance}


\author{Andrei Chernov}
\affiliation{%
  \institution{Independent Researcher}
  \country{Germany}}
\email{chernov.andrey.998@gmail.com}

\author{Haroon Wahab}
\affiliation{%
  \institution{School of Computer Science, AI and Electronics, University of Bradford}
  \country{United Kingdom}}
\email{m.h.wahab@bradford.ac.uk}

\author{Oleg Novitskij}
\affiliation{%
  \institution{Independent Researcher}
  \country{Russia}}
\email{o.a.novitskij@gmail.com}

\maketitle

\renewcommand{\shortauthors}{Chernov et al.}

\section*{Abstract}
 In recent years, various approaches have been proposed to leverage large language models (LLMs) for incorporating textual information about items into recommender systems. Existing methods primarily focus on either fine-tuning LLMs to generate recommendations or integrating LLM-based embeddings into downstream models. In this work, we follow the latter direction and propose \textbf{TextGCN}, which applies parameter-free graph convolution layers directly over LLM-based item-title embeddings, instead of learning ID-based embeddings as in traditional methods. By combining language semantics with graph message passing, this architecture achieves state-of-the-art zero-shot performance, significantly outperforming prior approaches. Furthermore, we introduce \textbf{TextGCN-MLP}, which extends TextGCN with a trainable multilayer perceptron trained using a contrastive loss, achieving state-of-the-art in-domain performance on recommendation benchmarks. However, the zero-shot performance of TextGCN-MLP remains lower than that of TextGCN, highlighting the trade-off between in-domain specialization and zero-shot generalization. We release our code on github at  \href{https://github.com/ChernovAndrey/TFCE}{github.com/ChernovAndrey/TFCE}






\section{Introduction}

In today's attention economy \cite{schuster2024attention}, recommender systems have become an essential tool for filtering vast digital content to match user preferences. Collaborative filtering, the core technique behind many recommender systems, leverages user–item interaction histories to rank items according to likely future preferences. Traditional methods used matrix factorization (MF) to learn low-rank latent factors \cite{MF2009-original, he2017fastmatrixfactorizationonline}, while neural models later introduced non-linear mappings to better capture user interaction behaviors \cite{NCF2017}.

Over the past decade, Graph Convolutional Networks (GCNs) have become dominant in capturing high-order structures and learning user and item embeddings~\cite{wang2019neural, sun2019multi}. Yet, these methods rely solely on user–item interactions to learn the embedding space, overlooking the abundant textual information available for items that could encode semantic and implicit collaborative signals within the language space. Some earlier works explored this direction, but they relied on primitive language representations~\cite{BoW_text_features_music_MF-2013, knowledgebasedMF-2016}.

With recent advances in Large Language Models (LLMs), language representations have significantly improved, enabling new opportunities to integrate semantic signals into recommendation models. Two main approaches have emerged: (i) fine-tuning LLMs for recommendation tasks~\cite{tennenholtz2023demystifying, liao2023llara, zhang2025collm}, and (ii) extracting item embeddings from LLMs during preprocessing~\cite{liu2024llm, sheng2024language}. The former enables more interpretable recommendations but comes with high computational costs and limited generalizability. The latter offers a simpler and more scalable alternative by enriching item representations via metadata, though this line of work remains less explored. One of the recent works in this direction, AlphaRec~\cite{sheng2024language}, projects LLM-based item embeddings into the collaborative space via a trainable MLP, followed by graph convolution layers, achieving state-of-the-art in-domain performance. The authors also demonstrate that AlphaRec can be pretrained on multiple datasets and achieve state-of-the-art zero-shot performance on unseen datasets. However, it remains unclear which components of the model primarily contribute to in-domain performance and which are responsible for generalization.

In this work, we present the \textbf{TextGCN-MLP} model, which introduces several modifications to AlphaRec to improve in-domain accuracy. The key architectural change lies in the order of the MLP and GCL layers: we apply graph convolution layers (GCLs) directly on top of the LLM-based item embeddings, followed by a multilayer perceptron (MLP). Additional improvements include adopting a two-tower MLP architecture, with separate networks for user and item embeddings, and employing a $K$-positive contrastive loss~\cite{KCL2020} to better leverage multiple user–item interactions during training. We show empirically that this design substantially enhances in-domain performance. Moreover, it allows us to decompose the model into two distinct components. The first part, \textbf{TextGCN}, is a training-free module that applies GCLs over frozen LLM embeddings and can be treated as an independent model. The second component is learnable MLP networks that refines these representations through contrastive learning. Surprisingly, even without any trainable parameters, TextGCN outperforms traditional recommender models such as LightGCN, MF~\cite{MF}, and MultVAE~\cite{VAE}, demonstrating that LLM-enriched textual representations of item titles provide a superior alternative to learned ID-based embeddings.

While TextGCN-MLP achieves state-of-the-art in-domain performance, TextGCN attains state-of-the-art zero-shot performance. This indicates that MLP-based contrastive training on top of collaborative embeddings may give better performance gains within that recommendation domain and dataset distribution, however it reduces the generalization capability of language-enriched collaborative embeddings of TextGCN. This observation aligns with the well-known foundation models' transfer learning trade-off between generalization and specialization ~\cite{lin2023speciality}. In this context, TextGCN embeddings find their role analogous to the RecSys foundation model embeddings (generalized representations) that can be further aligned for the downstream domain by training a domain specific MLP adapter as done in TextGCN-MLP. 

\noindent To summarize, this work makes the following key contributions:
\begin{itemize}
    \item We propose TextGCN, which applies parameter-free graph convolutional layers (GCLs) to item embeddings derived from a Large Language Model (LLM), achieving state-of-the-art performance in zero-shot recommendation tasks.

    \item We introduce TextGCN-MLP, a two-tower MLP trained with a $k$-positive contrastive loss on top of TextGCN, which achieves state-of-the-art in-domain performance across multiple standard benchmarks.
    
    \item We conduct a comprehensive ablation study demonstrating that most performance gains are attributable to the use of TextGCN embeddings, while architectural modifications such as the two-tower MLP and $k$-positive sampling contribute incrementally.
    
    \item We show that despite its strong in-domain performance, TextGCN-MLP performs worse than TextGCN in zero-shot settings. This highlights a well-known trade-off between specialization and generalization observed in fine-tuned foundation models.
\end{itemize}

\section{Related Work}
\subsection{GCNs for Recommendation}

Graph Convolutional Networks (GCNs) have gained popularity in collaborative filtering recommendation systems over the last five years, becoming a go-to solution for recommendation models. Early approaches, such as NGCF \cite{wang2019neural} and GCCF \cite{sun2019multi}, were adapted from node classification tasks~\cite{kipf2016semi}. Later, LightGCN \cite{he2020lightgcn} was introduced and demonstrated that simplifying GCNs to retain only neighborhood aggregation can lead to improved performance. TextGCN can be viewed as replacing the learnable ID-based embeddings in LightGCN~\cite{he2020lightgcn} with fixed, LLM-derived embeddings.

\subsection{Contrastive Learning}
Contrastive learning has emerged as a strong paradigm for representation learning in recommender systems. Bayesian Personalized Ranking (BPR) loss has long been used as a contrastive objective for recommender systems \cite{BPR2009, he2020lightgcn}. BPR optimizes for pairwise preferences by sampling triplets from the user–item interaction matrix, while enforcing that a positive pair (user–interacted item) scores higher than a negative pair (user–non-interacted item). More recently, the InfoNCE loss has gained popularity in recommendation, where it contrasts a single positive pair against a set of randomly sampled negatives \cite{SGL2021, bias-awareInfoNCE2022, xsimGCL2023l, sheng2024language}. InfoNCE has demonstrated performance gains over BPR across several domains, including recommender systems. However, recent advances in contrastive learning suggest that incorporating multiple positives per anchor yields more robust representations and leads to better alignment in the feature space \cite{KCL2020}.

In a typical collaborative filtering setting, multiple positives per user are naturally available. In this work, we propose using multiple positives per user to be sampled from interaction history, conducing stronger positive supervision signal per user in a mini-batch. We adopt the $k$-positive contrastive loss \cite{KCL2020} for recommender systems and investigate the impact of different values of $k$.

\subsection{LLMs in Recommendation Systems}
Despite the undeniable success of GCNs, they are typically based solely on the user-item implicit feedback matrix and do not incorporate any additional item features. With the emergence of Large Language Models (LLMs), studies aiming to integrate textual information about items, such as titles or descriptions, have started to appear. 

There are two main approaches to achieve this. The first relies on fine-tuning LLMs for use in recommendation systems \cite{tennenholtz2023demystifying, liao2023llara, zhang2025collm}. The second approach leverages LLMs to extract textual embeddings of items during preprocessing, without using LLMs during training or inference \cite{liu2024llm, sheng2024language}. The fine-tuning approach enables more human-interpretable recommendations by directly leveraging LLMs at prediction time. However, it requires significantly more computational resources at both training and inference stages. Furthermore, generalizing this approach to unseen datasets remains challenging. In contrast, using LLMs to enrich item representations, particularly through text-based metadata, is a simpler alternative, though it remains in its early stages. While we believe both directions have great potential, in this paper, we focus on the latter.

\subsection{Zero-Shot Learning}
The zero‑shot approach has been widely embraced since GPT models demonstrated impressive zero‑shot capabilities in natural language processing (NLP) \cite{Radford2019LanguageMA, GPT3, achiam2023gpt}. Such models are often referred to as foundation models in the literature. Foundation models are now emerging not only in NLP but also in computer vision \cite{radford2021learning}, time series forecasting \cite{ansari2024chronos}, and tabular data forecasting \cite{hollmann2022tabpfn}. In the field of recommendation systems, few models have been proposed for sequential zero-shot recommendation \cite{ding2021zero, hou2022towards}. In this paper, we focus on general (non-sequential) recommendation, where interactions are modeled without temporal order. For general recommenders, initial attempts at building foundation models \cite{P5,sheng2024language}, have been made, but their zero‑shot performance is not yet sufficient for broad adoption in recommendation tasks. In this paper, we take a step toward closing this gap in general recommendation by introducing TextGCN.

\section{Preliminaries}

\subsection{Collaborative Filtering}

In this paper, we focus on the collaborative filtering problem under implicit feedback settings using a binary interaction matrix. We model user preferences based on observed interactions, such as clicks or views, which are represented as binary signals. Namely, let $\mathcal{U} = \{u_1, u_2, \ldots, u_M\}$ be the set of users and $\mathcal{I} = \{i_1, i_2, \ldots, i_N\}$ the set of items. A binary interaction matrix $R \in \{0, 1\}^{M \times N}$, where $R_{ui} = 1$ indicates that user $u$ has interacted with item $i$, and $R_{ui} = 0$ otherwise. The goal of the collaborative filtering problem is to recommend, for each user, a set of items they have not yet interacted with but are likely to find relevant or be interested in.

\subsection{In-Domain vs Zero-Shot Evaluation}
In this paper, we evaluate the models’ performance in two different setups: in‑domain and zero‑shot.

\subsubsection{In-Domain Evaluation}
In‑domain evaluation refers to the traditional setting, where the model is applied to the same domain on which it was trained, and the test data largely share the same item set as the training data.

\subsubsection{Zero-Shot Evaluation}
Zero‑shot evaluation refers to the setting where a model is trained on one or several datasets but evaluated on datasets from unseen domains, meaning that the overlap between items in the training and test data is close to zero. This setting potentially enables the development of a universal model for a specific task that can perform well on previously unseen items.

\section{TextGCN}
\label{sec:TextGCN}
\begin{figure*}[t]
  \centering
  \includegraphics[width=\textwidth]{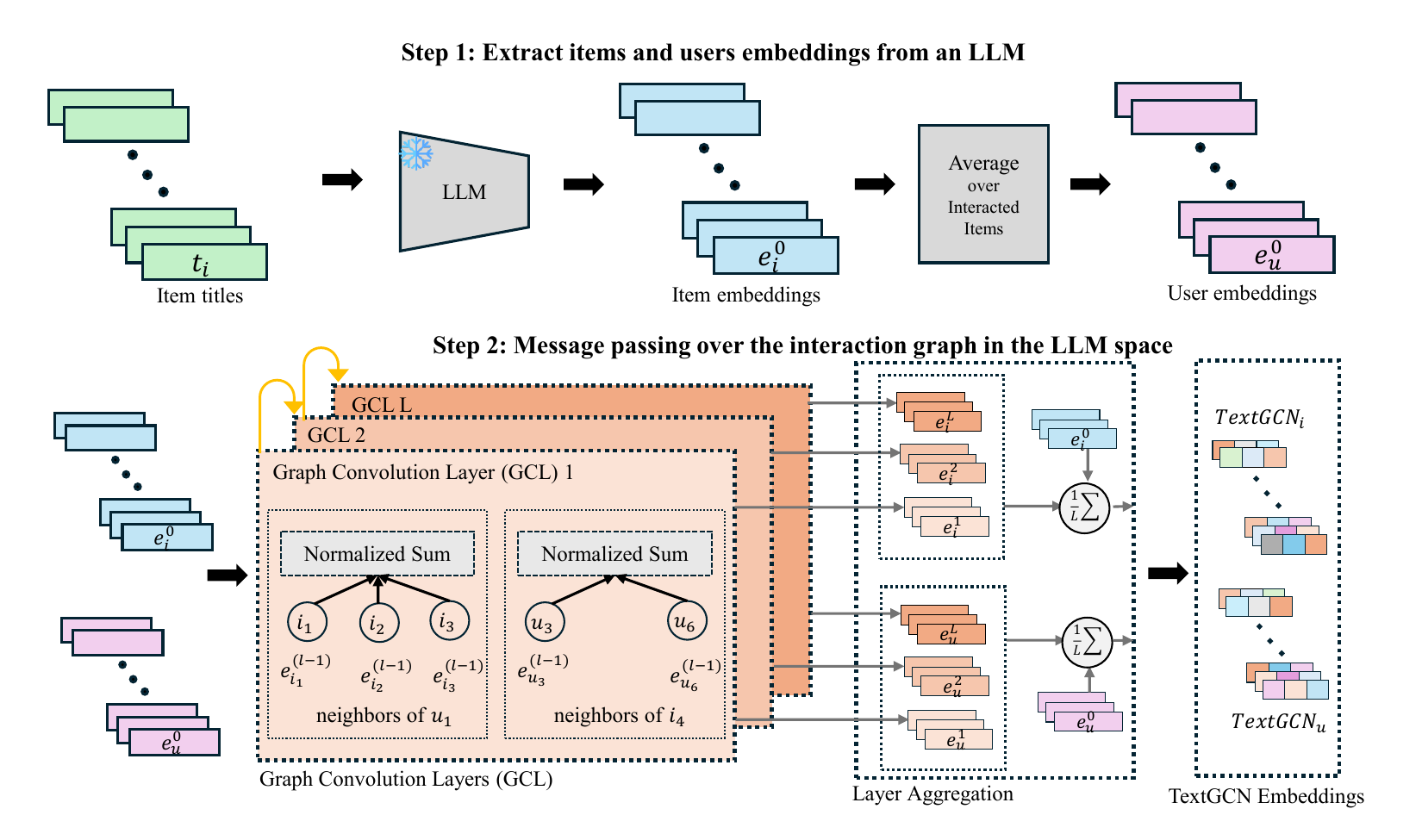}
  \caption{Proposed pipeline for obtaining training-free collaborative embeddings (TextGCN) in the language space. Modules with \textit{snowflake} are frozen (no trainable parameters).}
  \label{fig:TextGCN}

\end{figure*}

In this section, we present our main contribution: \textbf{TextGCN} Embeddings (TextGCN). TextGCN embeddings provide representations for both users and items. This approach requires no training, as it contains no learnable parameters. This makes it particularly suitable for zero-shot recommendation scenarios. We obtain TextGCN embeddings in two steps, which are described in the following sections (see Section~\ref{sec:TextGCN-llm} and Section~\ref{sec:TextGCN-gcl}) and illustrated in Figure~\ref{fig:TextGCN}.

To generate top-$k$ recommendations for a user using TextGCN, we compute the cosine similarity between the user's embedding and all non-interacted item embeddings. The items are then ranked based on their similarity scores, and the top-$k$ items with the highest values are recommended to the user.

\subsection{LLM Embeddings}
\label{sec:TextGCN-llm}
To make our approach applicable to a wide range of industrial datasets, we use only minimal textual information about items to extract embeddings.
Specifically, we generate embeddings solely from item titles to obtain LLM-based item representations, utilizing the \texttt{text-embedding-3-large} model \cite{neelakantan2022text}, and disregard any additional textual information available in the datasets. User embeddings are then computed as the average of the embeddings of all items the user interacted with in the training set.

\subsection{Diffusion over Interaction Graph}
\label{sec:TextGCN-gcl}
To incorporate collaborative structure into the learned representations, we construct a bipartite interaction graph $\mathcal{G} = (\mathcal{U} \cup \mathcal{I}, \mathcal{E})$, where an edge $(u, i) \in \mathcal{E}$ exists if $R_{ui} = 1$. We then apply $L$ graph convolutional layers (see Equation~\ref{eq:embedding_update}) to diffuse information across the graph, thereby enriching the LLM-derived embeddings with collaborative signals from user–item interactions.

\begin{equation}
\begin{aligned}
\mathbf{e}_u^{(l+1)} &= \sum_{i \in \mathcal{N}_u} \frac{1}{\sqrt{|\mathcal{N}_u|} \sqrt{|\mathcal{N}_i|}} \, \mathbf{e}_i^{(l)} \\
\mathbf{e}_i^{(l+1)} &= \sum_{u \in \mathcal{N}_i} \frac{1}{\sqrt{|\mathcal{N}_i|} \sqrt{|\mathcal{N}_u|}} \, \mathbf{e}_u^{(l)}
\end{aligned}
\label{eq:embedding_update}
\end{equation}

Finally, to obtain the TextGCN embeddings, we compute the average of the LLM embeddings (layer zero) and all subsequent layers produced by the graph convolution:

\begin{equation}
\mathbf{e}_u = \frac{1}{L+1} \sum_{l=0}^{L} \mathbf{e}_u^{(l)}, \quad
\mathbf{e}_i = \frac{1}{L+1} \sum_{l=0}^{L} \mathbf{e}_i^{(l)}.
\label{eq:final_embeddings}
\end{equation}

\section{TextGCN-MLP}
\label{sec:TextGCN-MLP}
\subsection{Architecture}

Once we obtain the TextGCN embeddings, we can train a neural network model on top of them to further improve performance. To validate this idea, we utilize an MLP block that takes TextGCN embeddings as input. Each MLP block consists of two linear layers with a LeakyReLU activation function~\cite{xu2020reluplex} applied between them. The input dimension of the MLP block is determined by the dimensionality of the LLM embeddings, which is $3072$. The hidden layer size is set to half of the input dimension, i.e., $1536$. The output dimension is treated as a tunable hyperparameter (see Section~\ref{sec:hyper_param} for details).

\begin{figure}[t]
  \centering
  \includegraphics[width=0.7\linewidth]{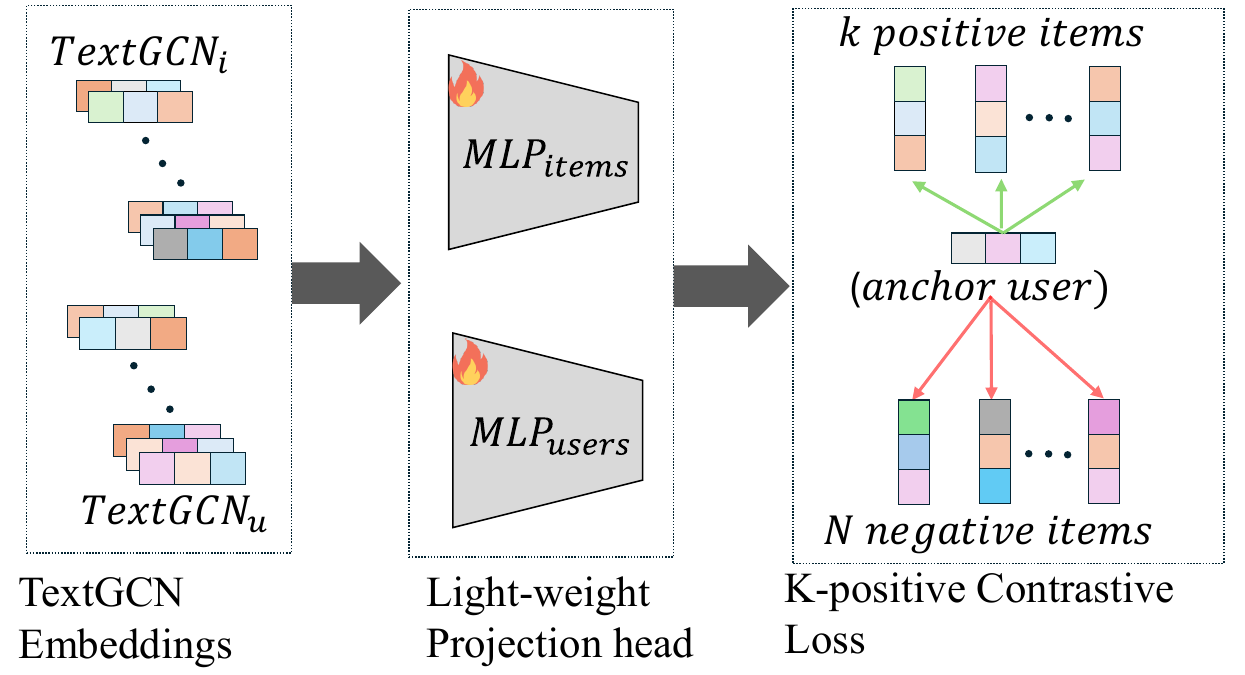}
  \caption{TextGCN-MLP training head using k positive contrastive loss (KCL). Modules with \textit{flame} allow gradient flow and parameter updates.}
  \label{fig:mlp_head}
\end{figure}

Finally, we employ two separate MLPs for user and item embeddings (see Figure~\ref{fig:mlp_head}), which share the same architecture but have independent weights. This design yields improved performance, as discussed in detail in Section~\ref{sec:ablation}.

\subsection{Contrastive Loss}
The contrastive objective is well-suited for modeling implicit collaborative filtering. To train the MLP blocks, we employ a contrastive loss function. InfoNCE has been predominantly adopted for collaborative filtering in recommender systems. However, the interaction matrix in collaborative filtering datasets is typically highly sparse, hence number of interacted items (positives) per user is often limited. This highlights the importance of fully utilizing all available positive signals within the contrastive learning framework. Motivated by this, we revisit the design of contrastive loss for collaborative filtering and adopt the $k$-positive contrastive loss (KCL) \cite{KCL2020}, which allows multiple positive samples per user to be leveraged effectively.

\begin{equation}
\mathcal{L}_{\text{KCL}} = \frac{1}{Nk} \sum_{u \in \mathcal{U}} \sum_{i^+ \in \mathcal{P}_k(u)} 
- \log \left( \frac{
    \exp\left(\text{sim}(e_u, e_{i^+}) / \tau\right)
}{Z_u(i^+)} \right),
\label{eq:kcl}
\end{equation}


\begin{equation}
\small
Z_u(i^+) = \exp\left(\frac{\text{sim}(e_u, e_{i^+})}{\tau}\right)
+ \sum_{j \in \mathcal{N}(u)} \exp\left(\frac{\text{sim}(e_u, e_j)}{\tau}\right)
\label{eq:partition}
\end{equation}

Where, $N$ denotes the number of users in the batch and $k$ is the number of positive items sampled per user. $\mathcal{U}$ is the set of users in the batch, and $\mathcal{P}_k(u) \subset \mathcal{P}(u)$ denotes $k$-positive items randomly sampled from user $u$'s interaction history. $\mathcal{N}(u)$ is the set of $J$ negative items for user $u$, sampled from the item pool excluding $\mathcal{P}(u)$. $e_u$ and $e_i$ represent the embeddings of user $u$ and item $i$, respectively. $\text{sim}(e_u, e_i)$ is a cosine similarity function, and $\tau$ is a temperature hyperparameter.

\section{Evaluation}
\label{sec:exp_setup}
\begin{table*}[ht]
\centering
\caption{\textbf{In-domain performance comparison.} TextGCN significantly outperforms all collaborative filtering (CF) baselines: MF, MultiVAE, and LightGCN, with statistical significance evaluated using the Recall metric ($p$-value $< 0.05$). TextGCN-MLP achieves the best overall performance, significantly surpassing the second-best method, AlphaRec ($p$-value $< 0.05$). In both cases, statistical significance was assessed using the Recall metric.}
\label{tab:main}
\resizebox{\linewidth}{!}{%
\begin{tabular}{llccc|ccc|ccc}
\toprule
& & \multicolumn{3}{c|}{Movies} & \multicolumn{3}{c|}{Games} & \multicolumn{3}{c}{Books} \\
& & Recall & NDCG & HR & Recall & NDCG & HR & Recall & NDCG & HR \\
\midrule
\multirow{6}{*}{\textbf{CF}} 
& MF & 0.0565 & 0.0515 & 0.3350 & 0.0327 & 0.0200 & 0.0883 & 0.0436 & 0.0389 & 0.2485 \\
& & {\scriptsize(± 0.0008)} & {\scriptsize(± 0.0007)} & {\scriptsize(± 0.0037)} & {\scriptsize(± 0.0005)} & {\scriptsize(± 0.0002)} & {\scriptsize(± 0.0011)} & {\scriptsize(± 0.0003)} & {\scriptsize(± 0.0002)} & {\scriptsize(± 0.0013)} \\ \addlinespace
& MultVAE & 0.0864 & 0.0784 & 0.4462 & 0.0908 & 0.0530 & 0.2094 & 0.0721 & 0.0598 & 0.3415 \\ 
& & {\scriptsize(± 0.0007)} & {\scriptsize(± 0.0009)} & {\scriptsize(± 0.0025)} & {\scriptsize(± 0.0006)} & {\scriptsize(± 0.0003)} & {\scriptsize(± 0.0011)} & {\scriptsize(± 0.0004)} & {\scriptsize(± 0.0004)} & {\scriptsize(± 0.0012)} \\ 
\addlinespace

& LightGCN & 0.0870 & 0.0766 & 0.4472 & 0.1025 & 0.0595 & 0.2327 & 0.0726 & 0.0609 & 0.3507 \\
& & {\scriptsize(± 0.0009)} & {\scriptsize(± 0.0008)} & {\scriptsize(± 0.0032)} & {\scriptsize(± 0.0004)} & {\scriptsize(± 0.0005)} & {\scriptsize(± 0.0009)} & {\scriptsize(± 0.0003)} & {\scriptsize(± 0.0002)} & {\scriptsize(± 0.0015)} \\ \addlinespace
\midrule
\multirow{3}{*}{\textbf{CL}} 
& SGL & 0.0916 & 0.0838 & 0.4680 & 0.1089 & 0.0634 & 0.2449 & 0.0789 & 0.0657 & 0.3734 \\
& BC Loss & 0.1039 & 0.0943 & 0.5037 & 0.1145 & 0.0668 & 0.2561 & 0.0915 & 0.0779 & 0.4045 \\
& XSimGCL & 0.1057 & 0.0984 & 0.5128 & 0.1138 & 0.0662 & 0.2550 & 0.0879 & 0.0745 & 0.3918 \\
\midrule
\multirow{4}{*}{\textbf{w/ LLM}} 
& KAR & 0.1084 & 0.1001 & 0.5134 & 0.1181 & 0.0693 & 0.2571 & 0.0852 & 0.0734 & 0.3834 \\
& RLMRec & 0.1119 & 0.1013 & 0.5301 & 0.1384 & 0.0809 & 0.2997 & 0.0928 & 0.0774 & 0.4092 \\
& EMB-KNN & 0.0548 & 0.0380 & 0.2916 & 0.0879 & 0.0389 & 0.1970 & 0.0434 & 0.0248 & 0.1851 \\
& AlphaRec & 0.1221 & 0.1138 & 0.5567 & 0.1511 & 0.0888 & 0.3188 & 0.0992 & 0.0832 & 0.4183 \\
& & {\scriptsize(± 0.0002)} & {\scriptsize(± 0.0003)} & {\scriptsize(± 0.0011)} & {\scriptsize(± 0.0006)} & {\scriptsize(± 0.0004)} & {\scriptsize(± 0.0012)} & {\scriptsize(± 0.0002)} & {\scriptsize(± 0.0002)} & {\scriptsize(± 0.0012)} \\
\addlinespace

\midrule
\multirow{4}{*}{\textbf{Ours}} 
& TextGCN & 0.0880 & 0.0814 & 0.4397 & 0.1319 & 0.0769 & 0.2828 & 0.0743 & 0.0627 & 0.4217 \\
& \textbf{TextGCN-MLP} & \textbf{0.1295} & \textbf{0.1227} & \textbf{0.5791} & \textbf{0.1581} & \textbf{0.0928} & \textbf{0.3315} & \textbf{0.1049} & \textbf{0.0875} & \textbf{0.4324} \\
& & {\scriptsize(± 0.0001)} & {\scriptsize(± 0.0007)} & {\scriptsize(± 0.0007)} & {\scriptsize(± 0.0005)} & {\scriptsize(± 0.0003)} & {\scriptsize(± 0.0012)} & {\scriptsize(± 0.0003)} & {\scriptsize(± 0.0002)} & {\scriptsize(± 0.0010)} \\
\addlinespace

\midrule
& Imp.\% over the best baseline & 6.55\% & 7.26\% & 4.21\% & 4.81\% & 4.70\% & 3.84\% & 6.56\% & 6.52\% & 3.82\% \\
\bottomrule
\end{tabular}
}
\end{table*}

\begin{table*}[ht]
\centering
\caption{\textbf{Zero-shot test performance comparison.} TextGCN substantially outperforms all baseline methods.}
\label{tab:zero-shot}
\resizebox{\textwidth}{!}{%
\begin{tabular}{llccc|ccc|ccc}
\toprule
& & \multicolumn{3}{c|}{Software} & \multicolumn{3}{c|}{Musical Instruments} & \multicolumn{3}{c}{Office Products} \\
& & Recall & NDCG & HR & Recall & NDCG & HR & Recall & NDCG & HR \\
\midrule
\multirow{4}{*}{\textbf{Baselines}} 
& Random & 0.0015 & 0.0006 & 0.0023 & 0.0011 & 0.0004 & 0.0019 & 0.0003 & 0.0001 & 0.0006 \\
& Pop & 0.0362 & 0.0141 & 0.0575 & 0.0131 & 0.0061 & 0.0283 & 0.0074 & 0.0029 & 0.0136 \\
& LightGCN (in-domain) & 0.0614 & 0.0250 & 0.0961 & 0.0177 & 0.0075 & 0.0332 & 0.0086 & 0.0038 & 0.0142 \\
& AlphaRec & 0.0687 & 0.0302 & 0.1002 & 0.0170 & 0.0073 & 0.0307 & 0.0113 & 0.0055 & 0.0181 \\
\midrule
\multirow{2}{*}{\textbf{Ours}} 
& TextGCN &  \textbf{0.0913} & \textbf{0.0408} & \textbf{0.1298}  & \textbf{0.0247} & \textbf{0.0111} & \textbf{0.0428} & \textbf{0.0162} & \textbf{0.0079} & \textbf{0.0254} \\
& TextGCN-MLP & 0.0724 & 0.0329 & 0.1044 & 0.0169 & 0.0075 & 0.0309 & 0.0088 & 0.0040 & 0.0135 \\
\midrule
& Imp.\% over the best baseline & 32.90\% & 35.10\% & 29.54\% & 39.5\% & 48.00\% & 28.92\% & 43.36\% & 43.64\% & 40.33\% \\
\bottomrule
\end{tabular}%
}
\end{table*}


\subsection{In-Domain}
\label{sec:in-domain}
To estimate in-domain performance, we use three Amazon 2018 datasets, which have become a standard choice for evaluating general-purpose recommender systems: \textit{Movies \& TV}, \textit{Video Games}, and \textit{Books}. Each dataset is split into training, validation, and test sets. The validation set is used for hyperparameter tuning (see Section~\ref{sec:hyper_param}) and early stopping during training, while the test set is reserved for final evaluation. We compare the performance of all models on their respective test sets.

\subsection{Zero-Shot}
\label{sec:zero-shot}
To estimate zero-shot performance, we jointly train on the training splits of the three Amazon 2018 datasets described in Section~\ref{sec:in-domain} and evaluate the resulting models on the test sets of three Amazon 2023 datasets: \textit{Software}, \textit{Musical Instruments}, and \textit{Office Products}. We use the latest available versions of the Amazon datasets, motivated by findings in \cite{LLMsMemorize}, which showed that LLMs are capable of memorizing publicly available datasets. Since we use the \texttt{text-embedding-3-large} model from OpenAI, whose knowledge cutoff is September 2021 \cite{openai_embeddings}, it is safe to assume that this model has not been exposed to the Amazon 2023 datasets.

\subsection{Metrics}
\label{sec:metrics}

We evaluate recommendation performance using three standard top-$k$ ranking metrics: Recall@k, which serves as our primary evaluation metric; NDCG@k (Normalized Discounted Cumulative Gain); and Hit Rate@k (HR@k). All metrics are computed per user and averaged across the test set. Unless stated otherwise, we report results at $k=20$.

Recall@k measures the proportion of relevant items retrieved among all relevant items for a given user. NDCG@k (Normalized Discounted Cumulative Gain) rewards relevant items that appear higher in the ranking by applying a logarithmic discount. HR@k is a binary metric that indicates whether at least one relevant item appears among the top-$k$ recommendations.

\subsection{Training}
\label{sec:training}
All models are trained until the Recall@$20$ metric on the validation set does not improve for $20$ consecutive epochs. We use the Adam optimizer for training and keep the learning rate constant throughout all training epochs. All evaluations were conducted on an NVIDIA V100 GPU with 16GB of memory, running on the CentOS 7 operating system.

\section{Experimental Setup}

\subsection{In-domain Baseline Models}

\subsubsection{Collaborative Filtering}
This group is referred to as \textit{CF} in Table~\ref{tab:main}. It includes classical collaborative filtering methods such as:
\begin{itemize}
    \item MF~\cite{MF} — A matrix factorization model.
    \item MultVAE~\cite{VAE} — A variational autoencoder-based model that generates item recommendations from a multinomial distribution.
    \item LightGCN~\cite{he2020lightgcn} — A graph-based model that learns embeddings and propagates them through graph convolution layers.
\end{itemize}

\subsubsection{Models with Contrastive Learning}
This group is denoted as \textit{CL} in Table~\ref{tab:main}, as all models in this category use contrastive loss during training. The models included are: SGL~\cite{SGL2021}, BC Loss~\cite{bias-awareInfoNCE2022}, and XSimGCL~\cite{xsimGCL2023l}.

\subsubsection{LLM-Enhanced Approaches}
This group consists of methods that utilize large language models (LLMs) and is labeled as \textit{w/ LLM} in Table~\ref{tab:main}. For a fair comparison, we use the same LLM-based representations in our model and in the baselines from this group. The models include:
\begin{itemize}
    \item KAR~\cite{KAR} — Combines generated textual information about users and items with traditional collaborative filtering using a hybrid expert adaptor.
    \item RLMRec~\cite{RLMRec} — Aligns the semantic space of LLMs with collaborative signals through cross-view alignment.
    \item EMB-KNN~\cite{sheng2024language} — A training-free method, that directly computes recommendations based on similarity between LLM embeddings. It can be seen as a special case of TextGCN with zero graph convolution layers.
    \item AlphaRec~\cite{sheng2024language} — Learns a projection function from language space to collaborative space. It shares similarities with our TextGCN-MLP model; we discuss this in more detail in Section~\ref{sec:diff-alpharec}.
\end{itemize}

\subsection{Zero-shot Baseline Models}
For zero-shot recommendation, we report performance on the test datasets for the following models:
\begin{itemize}
    \item Random: recommends items uniformly at random.
    \item Pop: recommends the most popular items from the training set to every user in the test set.
    \item LightGCN: trained independently on each of the three Amazon 2023 datasets in an in-domain setup.
    \item AlphaRec: trained jointly on all three Amazon 2018 datasets and evaluated on the Amazon 2023 datasets.
\end{itemize}

\subsection{Hyperparameters Tuning}
\label{sec:hyper_param}
For the baseline models, we use the optimal hyperparameter configurations reported in their respective papers.

For TextGCN-MLP, during in-domain experiments, we treat the output dimensionality of the MLP layers (\texttt{output\_size}), the number of graph convolutional layers (\texttt{n\_layers}), the temperature parameter for contrastive loss (\texttt{tau}), and the number of negative samples (\texttt{neg\_samples}) as tunable hyperparameters. These are optimized independently for each in-domain dataset using the validation Recall metric. 
For the number of positive samples (\texttt{pos\_samples}), we search over quantiles of the per-user item interaction distribution, and find that the median value performs best across all datasets. While increasing the number of negative samples beyond $512$ could potentially yield further improvements, it significantly increases both memory consumption and training time. The optimal parameter values and additional details on hyperparameter tuning are provided in Appendix~\ref{app:hyperparam}.

For the TextGCN model, there is only one hyperparameter: the number of graph convolutional layers (\texttt{n\_layers}). We tune it using the same approach as for TextGCN-MLP, and the optimal values are $4$, $3$, and $5$ for the \textit{Video Games}, \textit{Movies \& TV}, and \textit{Books} datasets, respectively. Yet, the performance gain from tuning TextGCN is almost negligible, as it achieves nearly the same results with the number of layers set to $2$.

To evaluate zero-shot performance, we do not tune any parameters on the zero-shot datasets, as this would undermine the dataset-agnostic objective of the experiment. For TextGCN, we set the number of graph convolutional layers to $2$. For the TextGCN-MLP model, we use the same set of hyperparameters as in the in-domain training on the \textit{Books} dataset.

\section{Results}
\label{sec:model_performance}
\subsection{In-domain Performance}
In Table~\ref{tab:main}, we report evaluation metrics on the test sets for the three in-domain datasets. We present results for our two models, TextGCN and TextGCN-MLP, as well as for all baseline models. For collaborative filtering (CF) baselines, AlphaRec, and our TextGCN-MLP model, we run experiments with five random seeds to assess the statistical significance of improvements. Specifically, we compare TextGCN against CF models, and TextGCN-MLP against AlphaRec, since AlphaRec consistently ranks second across all datasets with a clear margin over other baselines.

For models evaluated with five seeds, we report both the mean and standard deviation for each metric. Since TextGCN does not involve any learnable parameters and produces deterministic evaluation metrics, multiple seed runs are not needed.

\subsection{Zero-shot}
In Table~\ref{tab:zero-shot}, we present zero-shot performance on the three Amazon 2023 datasets for our models and the baselines. Since training TextGCN-MLP and AlphaRec on multiple datasets is time consuming, we report results from a single seed for these models. For the remaining zero-shot baselines, multiple seed runs are not required, as they do not involve any learnable parameters and produce deterministic outputs.

\subsection{Takeaways and Insights}

\noindent The results demonstrate the following:
\begin{itemize}
    \item TextGCN achieves significantly higher in-domain performance than traditional collaborative filtering models ($p$-value $< 0.05$).
    \item TextGCN-MLP achieves the highest in-domain results, significantly outperforming AlphaRec, the second-best model ($p$-value $< 0.05$).
    \item TextGCN achieves the highest zero-shot performance, outperforming all other models by a substantial margin.
\end{itemize}

In the ablation study (Section~\ref{sec:ablation}), we show that incorporating a two-tower MLP and $k$-positive contrastive loss contributes only marginally to the overall performance of the TextGCN-MLP model. This suggests that most performance gains stem from domain adaptation achieved by learning an MLP layer on top of the TextGCN output.

However, this domain adaptation comes at a cost. On zero-shot datasets, TextGCN-MLP and AlphaRec perform comparably, reaching levels similar to LightGCN’s in-domain performance, yet both fall substantially short of TextGCN. These results indicate that MLP layers enhance domain-specific adaptation but do not improve generalization across datasets, instead leading to specialization that ultimately reduces cross-domain robustness.

\section{Discussion}
In this section, we show that applying graph convolutional layers (GCLs) over LLM-based embeddings is a key component of the TextGCN-MLP architecture (see Section~\ref{sec:ablation}). Additionally, we compare the architectures of TextGCN-MLP and AlphaRec (see Section~\ref{sec:diff-alpharec}).

\subsection{Ablation Study for TextGCN-MLP}
\label{sec:ablation}

The TextGCN-MLP model consists of three key architectural components:
\begin{itemize}
    \item Applying graph convolution layers (GCL) over LLM-based item embeddings during preprocessing, instead of learning ID-based embeddings.
    \item Using two separate MLPs, one for user embeddings and one for item embeddings.
    \item Training with a $k$-positive contrastive loss, where $k$ is set to the median number of item interactions per user.
\end{itemize}

Here, we investigate the individual contributions of each component. While the second and third components provide incremental improvements, we find that the first component, applying GCLs over LLM-based item embeddings, contributes the most to the overall performance. Notably, TextGCN-MLP outperforms all baselines even when using a single MLP and standard InfoNCE loss with one positive sample.

To illustrate the contribution of each component, we compare four ablated variants of the \textbf{TextGCN-MLP} model:
\begin{itemize}
    \item \textit{One Tower / 1-pos}: A single MLP is shared between user and item embeddings. The model is trained using standard InfoNCE loss with one positive sample.
    \item \textit{One Tower / $k$-pos}: A single MLP is shared, but training employs a $k$-positive contrastive loss, where $k$ is set to the median number of user-item interactions per user.
    \item \textit{Two-tower / 1-pos}: Two separate MLPs are used for user and item embeddings. Training still uses a single positive sample per anchor.
    \item \textit{Two-tower / $k$-pos}: The full version of \textbf{TextGCN-MLP} used in all experiments, incorporating separate MLPs for user and item embeddings and training with $k$-positive contrastive loss.
\end{itemize}

In-domain test Recall scores for each configuration are reported in Table~\ref{tab:ablation_textgcn_mlp}. While employing a two-tower architecture and $k$-positive contrastive loss further improves performance, the \textit{One Tower / 1-pos} variant of TextGCN-MLP already surpasses all baseline models.

\begin{table}[h]
\centering
\caption{Recall on the test sets for four variants of the TextGCN-MLP model. See Section~\ref{sec:ablation} for details.} 
\label{tab:ablation_textgcn_mlp}
\begin{tabular}{lccc}
\toprule
\textbf{Model} & \textbf{Movies} & \textbf{Games} & \textbf{Books} \\
\midrule
One Tower / 1-pos    & 0.1281 & 0.1556 & 0.1032 \\
\midrule
One Tower / $k$-pos    & \underline{0.1291} & 0.1567 & 0.1044 \\
\midrule
Two-tower / 1-pos    & 0.1269 & \underline{0.1577} & \underline{0.1045} \\
\midrule
Two-tower / $k$-pos & \textbf{0.1295} & \textbf{0.1581} & \textbf{0.1049} \\
\bottomrule
\end{tabular}
\end{table}

\subsection{TextGCN-MLP vs. AlphaRec}
\label{sec:diff-alpharec}

\subsection{Architecture}

\begin{table}[h]
\centering
\caption{Recall on the test sets for four variants of AlphaRec baseline model. See Section~\ref{sec:diff-alpharec} for details.} 
\label{tab:two_tower}
\label{tab:ablation_textgcn_mlp}
\begin{tabular}{lccc}
\toprule
\textbf{Model} & \textbf{Movies} & \textbf{Games} & \textbf{Books} \\
\midrule
One Tower / 1-pos    & \textbf{0.1221} & \underline{0.1511} & 0.0992 \\
\midrule
One Tower / $k$-pos    & 0.1211 & \textbf{0.1526} & \underline{0.0998} \\
\midrule
Two-tower / 1-pos    & 0.1207 & 0.1501 & 0.0988 \\
\midrule
Two-tower / $k$-pos    & \underline{0.1214} & \underline{0.1511} & \textbf{0.1006} \\
\bottomrule
\end{tabular}
\end{table}

\begin{figure}[t]
  \centering
  \includegraphics[width=0.7\linewidth]{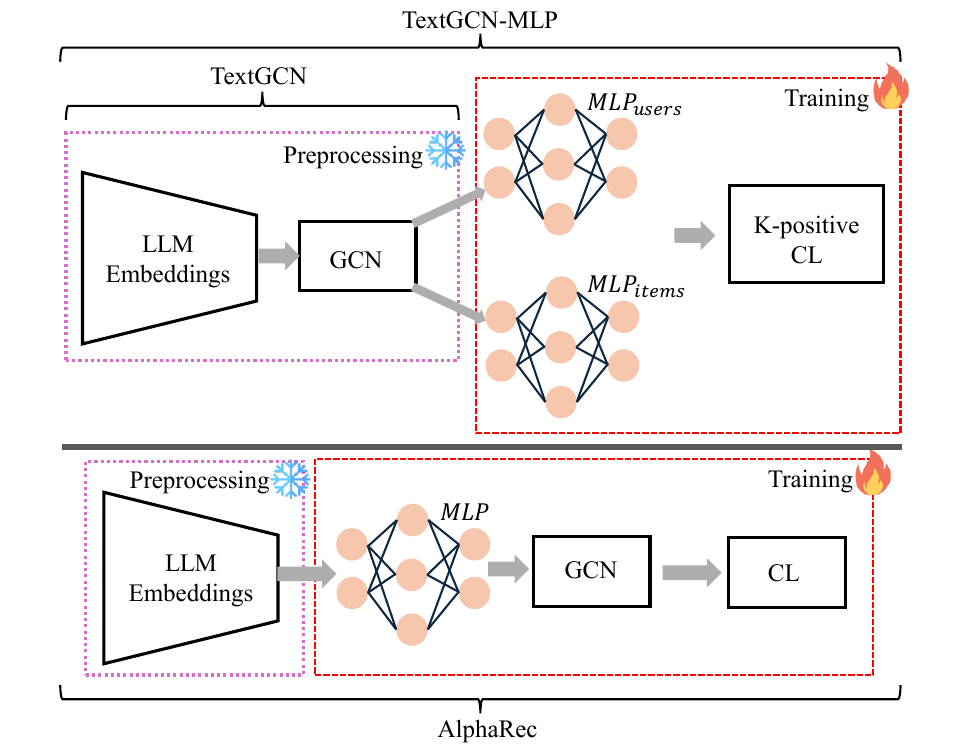}
  \caption{Comparison of TextGCN-MLP with AlphaRec architecture}
  \label{fig:comp_alpha_textgcn}
\end{figure}

The main distinction between AlphaRec and TextGCN-MLP is that TextGCN-MLP operates on top of fixed TextGCN embeddings, whereas AlphaRec learns a direct alignment between the language and collaborative spaces. AlphaRec first projects LLM-based item embeddings into the collaborative space using a learnable MLP, and then applies graph convolution layers (GCL) (see Figure~\ref{fig:comp_alpha_textgcn}).

As a result, unlike TextGCN-MLP, AlphaRec must propagate gradients through the GCL layers during training, which has been shown to impair gradient flow and hinder optimization \cite{GCLhurts1,GCLhurts2}. Consequently, TextGCN-MLP exhibits more stable training compared to AlphaRec. This also explains why architectural enhancements such as a two-tower MLP or $k$-positive contrastive loss do not lead to performance improvements in AlphaRec (see Table~\ref{tab:ablation_textgcn_mlp}).

\subsection{Efficiency}
We compare the computational efficiency of TextGCN-MLP and AlphaRec on the in-domain datasets. Tables~\ref{tab:train-time} and~\ref{tab:test-time} present the training and evaluation time comparisons, respectively.

We observe that TextGCN-MLP achieves faster per-iteration performance in both training and evaluation. This efficiency stems from applying the graph convolution only once during preprocessing, whereas AlphaRec performs multiple graph convolution layers at every forward pass. However, AlphaRec converges faster during training.

\begin{table}[h]
\centering
\caption{Training time comparison (seconds per epoch / total).}
\label{tab:train-time}
\begin{adjustbox}{max width=\linewidth}
\begin{tabular}{l|ccc}
\toprule
\textbf{Model} & \textbf{Video Games} & \textbf{Movies \& TV} & \textbf{Books} \\
\midrule
AlphaRec            & 12.8 / 627.1 & 13.4 / 790.98 & 74.7 / 4932.6 \\
TextGCN-MLP  & 11.5 / 1243.4 & 11.8 / 1520.8 & 71.3 / 8410.9 \\
\bottomrule
\end{tabular}
\end{adjustbox}
\end{table}

\begin{table}[h]

\centering
\caption{Evaluation time comparison (seconds on both validation and test datasets combined).}
\label{tab:test-time}
\begin{adjustbox}{max width=\linewidth}
\begin{tabular}{l|ccc}
\toprule
\textbf{Model} & \textbf{Video Games} & \textbf{Movies \& TV} & \textbf{Books} \\
\midrule
AlphaRec            & 7.3  & 4.3  & 27.4 \\
TextGCN-MLP  & 6.3  & 3.5  & 26.7 \\
\bottomrule
\end{tabular}
\end{adjustbox}
\end{table}

\section{Conclusion}

In this work, we introduced TextGCN, a method that applies graph convolution directly to item embeddings derived from a large language model (LLM), without requiring any learnable parameters. As a result, TextGCN achieves efficient and state-of-the-art performance in zero-shot recommendation tasks.

To achieve strong in-domain performance, we extended this approach with a two-tower MLP architecture consisting of one MLP for user embeddings and another for item embeddings. The resulting model, TextGCN-MLP, is trained using a $k$-positive contrastive loss and achieves state-of-the-art in-domain performance across multiple datasets. Through an ablation study, we empirically demonstrated that although the two-tower architecture and $k$ positive contrastive loss contribute to performance, the primary gains come from the use of TextGCN embeddings. As a result, the TextGCN-MLP variant with a single tower MLP and a single positive sample in the contrastive loss still outperforms all baseline models.

However, the zero-shot performance of TextGCN-MLP remains lower than that of TextGCN, suggesting that MLPs tend to specialize in fitting dataset-specific patterns at the cost of generalization.

\bibliographystyle{ACM-Reference-Format}
\bibliography{sample-base}


\begin{thebibliography}{41}


\ifx \showCODEN    \undefined \def \showCODEN     #1{\unskip}     \fi
\ifx \showISBNx    \undefined \def \showISBNx     #1{\unskip}     \fi
\ifx \showISBNxiii \undefined \def \showISBNxiii  #1{\unskip}     \fi
\ifx \showISSN     \undefined \def \showISSN      #1{\unskip}     \fi
\ifx \showLCCN     \undefined \def \showLCCN      #1{\unskip}     \fi
\ifx \shownote     \undefined \def \shownote      #1{#1}          \fi
\ifx \showarticletitle \undefined \def \showarticletitle #1{#1}   \fi
\ifx \showURL      \undefined \def \showURL       {\relax}        \fi
\providecommand\bibfield[2]{#2}
\providecommand\bibinfo[2]{#2}
\providecommand\natexlab[1]{#1}
\providecommand\showeprint[2][]{arXiv:#2}

\bibitem[Achiam et~al\mbox{.}(2023)]%
        {achiam2023gpt}
\bibfield{author}{\bibinfo{person}{Josh Achiam}, \bibinfo{person}{Steven Adler}, \bibinfo{person}{Sandhini Agarwal}, \bibinfo{person}{Lama Ahmad}, \bibinfo{person}{Ilge Akkaya}, \bibinfo{person}{Florencia~Leoni Aleman}, \bibinfo{person}{Diogo Almeida}, \bibinfo{person}{Janko Altenschmidt}, \bibinfo{person}{Sam Altman}, \bibinfo{person}{Shyamal Anadkat}, {et~al\mbox{.}}} \bibinfo{year}{2023}\natexlab{}.
\newblock \showarticletitle{Gpt-4 technical report}.
\newblock \bibinfo{journal}{\emph{arXiv preprint arXiv:2303.08774}} (\bibinfo{year}{2023}).
\newblock


\bibitem[Ansari et~al\mbox{.}(2024)]%
        {ansari2024chronos}
\bibfield{author}{\bibinfo{person}{Abdul~Fatir Ansari}, \bibinfo{person}{Lorenzo Stella}, \bibinfo{person}{Caner Turkmen}, \bibinfo{person}{Xiyuan Zhang}, \bibinfo{person}{Pedro Mercado}, \bibinfo{person}{Huibin Shen}, \bibinfo{person}{Oleksandr Shchur}, \bibinfo{person}{Syama~Sundar Rangapuram}, \bibinfo{person}{Sebastian~Pineda Arango}, \bibinfo{person}{Shubham Kapoor}, {et~al\mbox{.}}} \bibinfo{year}{2024}\natexlab{}.
\newblock \showarticletitle{Chronos: Learning the language of time series}.
\newblock \bibinfo{journal}{\emph{arXiv preprint arXiv:2403.07815}} (\bibinfo{year}{2024}).
\newblock


\bibitem[Brown et~al\mbox{.}(2020)]%
        {GPT3}
\bibfield{author}{\bibinfo{person}{Tom~B. Brown}, \bibinfo{person}{Benjamin Mann}, \bibinfo{person}{Nick Ryder}, \bibinfo{person}{Melanie Subbiah}, \bibinfo{person}{Jared Kaplan}, \bibinfo{person}{Prafulla Dhariwal}, \bibinfo{person}{Arvind Neelakantan}, \bibinfo{person}{Pranav Shyam}, \bibinfo{person}{Girish Sastry}, \bibinfo{person}{Amanda Askell}, \bibinfo{person}{Sandhini Agarwal}, \bibinfo{person}{Ariel Herbert-Voss}, \bibinfo{person}{Gretchen Krueger}, \bibinfo{person}{Tom Henighan}, \bibinfo{person}{Rewon Child}, \bibinfo{person}{Aditya Ramesh}, \bibinfo{person}{Daniel~M. Ziegler}, \bibinfo{person}{Jeffrey Wu}, \bibinfo{person}{Clemens Winter}, \bibinfo{person}{Christopher Hesse}, \bibinfo{person}{Mark Chen}, \bibinfo{person}{Eric Sigler}, \bibinfo{person}{Mateusz Litwin}, \bibinfo{person}{Scott Gray}, \bibinfo{person}{Benjamin Chess}, \bibinfo{person}{Jack Clark}, \bibinfo{person}{Christopher Berner}, \bibinfo{person}{Sam McCandlish}, \bibinfo{person}{Alec Radford}, \bibinfo{person}{Ilya Sutskever},
  {and} \bibinfo{person}{Dario Amodei}.} \bibinfo{year}{2020}\natexlab{}.
\newblock \showarticletitle{Language models are few-shot learners}. In \bibinfo{booktitle}{\emph{Proceedings of the 34th International Conference on Neural Information Processing Systems}} (Vancouver, BC, Canada) \emph{(\bibinfo{series}{NIPS '20})}. \bibinfo{publisher}{Curran Associates Inc.}, \bibinfo{address}{Red Hook, NY, USA}, Article \bibinfo{articleno}{159}, \bibinfo{numpages}{25}~pages.
\newblock
\showISBNx{9781713829546}


\bibitem[Di~Palma et~al\mbox{.}(2025)]%
        {LLMsMemorize}
\bibfield{author}{\bibinfo{person}{Dario Di~Palma}, \bibinfo{person}{Felice~Antonio Merra}, \bibinfo{person}{Maurizio Sfilio}, \bibinfo{person}{Vito~Walter Anelli}, \bibinfo{person}{Fedelucio Narducci}, {and} \bibinfo{person}{Tommaso Di~Noia}.} \bibinfo{year}{2025}\natexlab{}.
\newblock \showarticletitle{Do LLMs Memorize Recommendation Datasets? A Preliminary Study on MovieLens-1M}. In \bibinfo{booktitle}{\emph{Proceedings of the 48th International ACM SIGIR Conference on Research and Development in Information Retrieval}} (Padua, Italy) \emph{(\bibinfo{series}{SIGIR '25})}. \bibinfo{publisher}{Association for Computing Machinery}, \bibinfo{address}{New York, NY, USA}, \bibinfo{pages}{2582–2586}.
\newblock
\showISBNx{9798400715921}
\href{https://doi.org/10.1145/3726302.3730178}{doi:\nolinkurl{10.1145/3726302.3730178}}


\bibitem[Ding et~al\mbox{.}(2021)]%
        {ding2021zero}
\bibfield{author}{\bibinfo{person}{Hao Ding}, \bibinfo{person}{Yifei Ma}, \bibinfo{person}{Anoop Deoras}, \bibinfo{person}{Yuyang Wang}, {and} \bibinfo{person}{Hao Wang}.} \bibinfo{year}{2021}\natexlab{}.
\newblock \showarticletitle{Zero-shot recommender systems}.
\newblock \bibinfo{journal}{\emph{arXiv preprint arXiv:2105.08318}} (\bibinfo{year}{2021}).
\newblock


\bibitem[Geng et~al\mbox{.}(2022)]%
        {P5}
\bibfield{author}{\bibinfo{person}{Shijie Geng}, \bibinfo{person}{Shuchang Liu}, \bibinfo{person}{Zuohui Fu}, \bibinfo{person}{Yingqiang Ge}, {and} \bibinfo{person}{Yongfeng Zhang}.} \bibinfo{year}{2022}\natexlab{}.
\newblock \showarticletitle{Recommendation as Language Processing (RLP): A Unified Pretrain, Personalized Prompt \& Predict Paradigm (P5)}. In \bibinfo{booktitle}{\emph{Proceedings of the 16th ACM Conference on Recommender Systems}} (Seattle, WA, USA) \emph{(\bibinfo{series}{RecSys '22})}. \bibinfo{publisher}{Association for Computing Machinery}, \bibinfo{address}{New York, NY, USA}, \bibinfo{pages}{299–315}.
\newblock
\showISBNx{9781450392785}
\href{https://doi.org/10.1145/3523227.3546767}{doi:\nolinkurl{10.1145/3523227.3546767}}


\bibitem[He et~al\mbox{.}(2020)]%
        {he2020lightgcn}
\bibfield{author}{\bibinfo{person}{Xiangnan He}, \bibinfo{person}{Kuan Deng}, \bibinfo{person}{Xiang Wang}, \bibinfo{person}{Yan Li}, \bibinfo{person}{Yongdong Zhang}, {and} \bibinfo{person}{Meng Wang}.} \bibinfo{year}{2020}\natexlab{}.
\newblock \showarticletitle{Lightgcn: Simplifying and powering graph convolution network for recommendation}. In \bibinfo{booktitle}{\emph{Proceedings of the 43rd International ACM SIGIR conference on research and development in Information Retrieval}}. \bibinfo{pages}{639--648}.
\newblock


\bibitem[He et~al\mbox{.}(2017a)]%
        {NCF2017}
\bibfield{author}{\bibinfo{person}{Xiangnan He}, \bibinfo{person}{Lizi Liao}, \bibinfo{person}{Hanwang Zhang}, \bibinfo{person}{Liqiang Nie}, \bibinfo{person}{Xia Hu}, {and} \bibinfo{person}{Tat-Seng Chua}.} \bibinfo{year}{2017}\natexlab{a}.
\newblock \showarticletitle{Neural Collaborative Filtering}. In \bibinfo{booktitle}{\emph{Proceedings of the 26th International Conference on World Wide Web}}. \bibinfo{pages}{173–182}.
\newblock


\bibitem[He et~al\mbox{.}(2017b)]%
        {he2017fastmatrixfactorizationonline}
\bibfield{author}{\bibinfo{person}{Xiangnan He}, \bibinfo{person}{Hanwang Zhang}, \bibinfo{person}{Min-Yen Kan}, {and} \bibinfo{person}{Tat-Seng Chua}.} \bibinfo{year}{2017}\natexlab{b}.
\newblock \bibinfo{title}{Fast Matrix Factorization for Online Recommendation with Implicit Feedback}.
\newblock
\urldef\tempurl%
\url{https://arxiv.org/abs/1708.05024}
\showURL{%
\tempurl}


\bibitem[Hollmann et~al\mbox{.}(2022)]%
        {hollmann2022tabpfn}
\bibfield{author}{\bibinfo{person}{Noah Hollmann}, \bibinfo{person}{Samuel M{\"u}ller}, \bibinfo{person}{Katharina Eggensperger}, {and} \bibinfo{person}{Frank Hutter}.} \bibinfo{year}{2022}\natexlab{}.
\newblock \showarticletitle{Tabpfn: A transformer that solves small tabular classification problems in a second}.
\newblock \bibinfo{journal}{\emph{arXiv preprint arXiv:2207.01848}} (\bibinfo{year}{2022}).
\newblock


\bibitem[Hou et~al\mbox{.}(2022)]%
        {hou2022towards}
\bibfield{author}{\bibinfo{person}{Yupeng Hou}, \bibinfo{person}{Shanlei Mu}, \bibinfo{person}{Wayne~Xin Zhao}, \bibinfo{person}{Yaliang Li}, \bibinfo{person}{Bolin Ding}, {and} \bibinfo{person}{Ji-Rong Wen}.} \bibinfo{year}{2022}\natexlab{}.
\newblock \showarticletitle{Towards universal sequence representation learning for recommender systems}. In \bibinfo{booktitle}{\emph{Proceedings of the 28th ACM SIGKDD conference on knowledge discovery and data mining}}. \bibinfo{pages}{585--593}.
\newblock


\bibitem[Jaiswal et~al\mbox{.}(2022)]%
        {GCLhurts1}
\bibfield{author}{\bibinfo{person}{Ajay Jaiswal}, \bibinfo{person}{Peihao Wang}, \bibinfo{person}{Tianlong Chen}, \bibinfo{person}{Justin Rousseau}, \bibinfo{person}{Ying Ding}, {and} \bibinfo{person}{Zhangyang Wang}.} \bibinfo{year}{2022}\natexlab{}.
\newblock \showarticletitle{Old can be gold: Better gradient flow can make vanilla-gcns great again}.
\newblock \bibinfo{journal}{\emph{Advances in Neural Information Processing Systems}}  \bibinfo{volume}{35} (\bibinfo{year}{2022}), \bibinfo{pages}{7561--7574}.
\newblock


\bibitem[Kang et~al\mbox{.}(2020)]%
        {KCL2020}
\bibfield{author}{\bibinfo{person}{Bingyi Kang}, \bibinfo{person}{Yu Li}, \bibinfo{person}{Sa Xie}, \bibinfo{person}{Zehuan Yuan}, {and} \bibinfo{person}{Jiashi Feng}.} \bibinfo{year}{2020}\natexlab{}.
\newblock \showarticletitle{Exploring balanced feature spaces for representation learning}. In \bibinfo{booktitle}{\emph{International conference on learning representations}}.
\newblock


\bibitem[Kipf(2016)]%
        {kipf2016semi}
\bibfield{author}{\bibinfo{person}{TN Kipf}.} \bibinfo{year}{2016}\natexlab{}.
\newblock \showarticletitle{Semi-Supervised Classification with Graph Convolutional Networks}.
\newblock \bibinfo{journal}{\emph{arXiv preprint arXiv:1609.02907}} (\bibinfo{year}{2016}).
\newblock


\bibitem[Koren et~al\mbox{.}(2009)]%
        {MF2009-original}
\bibfield{author}{\bibinfo{person}{Yehuda Koren}, \bibinfo{person}{Robert Bell}, {and} \bibinfo{person}{Chris Volinsky}.} \bibinfo{year}{2009}\natexlab{}.
\newblock \showarticletitle{Matrix Factorization Techniques for Recommender Systems}.
\newblock \bibinfo{journal}{\emph{Computer}} (\bibinfo{year}{2009}), \bibinfo{pages}{30--37}.
\newblock


\bibitem[Liang et~al\mbox{.}(2018)]%
        {VAE}
\bibfield{author}{\bibinfo{person}{Dawen Liang}, \bibinfo{person}{Rahul~G Krishnan}, \bibinfo{person}{Matthew~D Hoffman}, {and} \bibinfo{person}{Tony Jebara}.} \bibinfo{year}{2018}\natexlab{}.
\newblock \showarticletitle{Variational autoencoders for collaborative filtering}. In \bibinfo{booktitle}{\emph{Proceedings of the 2018 world wide web conference}}. \bibinfo{pages}{689--698}.
\newblock


\bibitem[Liao et~al\mbox{.}(2023)]%
        {liao2023llara}
\bibfield{author}{\bibinfo{person}{Jiayi Liao}, \bibinfo{person}{Sihang Li}, \bibinfo{person}{Zhengyi Yang}, \bibinfo{person}{Jiancan Wu}, \bibinfo{person}{Yancheng Yuan}, {and} \bibinfo{person}{Xiang Wang}.} \bibinfo{year}{2023}\natexlab{}.
\newblock \showarticletitle{Llara: Aligning large language models with sequential recommenders}.
\newblock \bibinfo{journal}{\emph{CoRR}} (\bibinfo{year}{2023}).
\newblock


\bibitem[Lin et~al\mbox{.}(2023)]%
        {lin2023speciality}
\bibfield{author}{\bibinfo{person}{Yong Lin}, \bibinfo{person}{Lu Tan}, \bibinfo{person}{Hangyu Lin}, \bibinfo{person}{Zeming Zheng}, \bibinfo{person}{Renjie Pi}, \bibinfo{person}{Jipeng Zhang}, \bibinfo{person}{Shizhe Diao}, \bibinfo{person}{Haoxiang Wang}, \bibinfo{person}{Han Zhao}, \bibinfo{person}{Yuan Yao}, {et~al\mbox{.}}} \bibinfo{year}{2023}\natexlab{}.
\newblock \showarticletitle{Speciality vs generality: An empirical study on catastrophic forgetting in fine-tuning foundation models}.
\newblock \bibinfo{journal}{\emph{arXiv preprint arXiv:2309.06256}}  \bibinfo{volume}{11} (\bibinfo{year}{2023}), \bibinfo{pages}{14}.
\newblock


\bibitem[Liu et~al\mbox{.}(2024)]%
        {liu2024llm}
\bibfield{author}{\bibinfo{person}{Qidong Liu}, \bibinfo{person}{Xian Wu}, \bibinfo{person}{Yejing Wang}, \bibinfo{person}{Zijian Zhang}, \bibinfo{person}{Feng Tian}, \bibinfo{person}{Yefeng Zheng}, {and} \bibinfo{person}{Xiangyu Zhao}.} \bibinfo{year}{2024}\natexlab{}.
\newblock \showarticletitle{Llm-esr: Large language models enhancement for long-tailed sequential recommendation}.
\newblock \bibinfo{journal}{\emph{Advances in Neural Information Processing Systems}}  \bibinfo{volume}{37} (\bibinfo{year}{2024}), \bibinfo{pages}{26701--26727}.
\newblock


\bibitem[Lukovnikov and Fischer(2021)]%
        {GCLhurts2}
\bibfield{author}{\bibinfo{person}{Denis Lukovnikov} {and} \bibinfo{person}{Asja Fischer}.} \bibinfo{year}{2021}\natexlab{}.
\newblock \showarticletitle{Improving Breadth-Wise Backpropagation in Graph Neural Networks Helps Learning Long-Range Dependencies.}. In \bibinfo{booktitle}{\emph{Proceedings of the 38th International Conference on Machine Learning}} \emph{(\bibinfo{series}{Proceedings of Machine Learning Research}, Vol.~\bibinfo{volume}{139})}, \bibfield{editor}{\bibinfo{person}{Marina Meila} {and} \bibinfo{person}{Tong Zhang}} (Eds.). \bibinfo{publisher}{PMLR}, \bibinfo{pages}{7180--7191}.
\newblock
\urldef\tempurl%
\url{https://proceedings.mlr.press/v139/lukovnikov21a.html}
\showURL{%
\tempurl}


\bibitem[Neelakantan et~al\mbox{.}(2022)]%
        {neelakantan2022text}
\bibfield{author}{\bibinfo{person}{Arvind Neelakantan}, \bibinfo{person}{Tao Xu}, \bibinfo{person}{Raul Puri}, \bibinfo{person}{Alec Radford}, \bibinfo{person}{Jesse~Michael Han}, \bibinfo{person}{Jerry Tworek}, \bibinfo{person}{Qiming Yuan}, \bibinfo{person}{Nikolas Tezak}, \bibinfo{person}{Jong~Wook Kim}, \bibinfo{person}{Chris Hallacy}, {et~al\mbox{.}}} \bibinfo{year}{2022}\natexlab{}.
\newblock \showarticletitle{Text and code embeddings by contrastive pre-training}.
\newblock \bibinfo{journal}{\emph{arXiv preprint arXiv:2201.10005}} (\bibinfo{year}{2022}).
\newblock


\bibitem[{OpenAI}(2023)]%
        {openai_embeddings}
\bibfield{author}{\bibinfo{person}{{OpenAI}}.} \bibinfo{year}{2023}\natexlab{}.
\newblock \bibinfo{title}{OpenAI text‑embedding‑3‑large Model Documentation}.
\newblock \bibinfo{howpublished}{\url{https://platform.openai.com/docs/guides/embeddings}}.
\newblock
\newblock
\shownote{Accessed: 2025‑07‑29}.


\bibitem[Radford et~al\mbox{.}(2021)]%
        {radford2021learning}
\bibfield{author}{\bibinfo{person}{Alec Radford}, \bibinfo{person}{Jong~Wook Kim}, \bibinfo{person}{Chris Hallacy}, \bibinfo{person}{Aditya Ramesh}, \bibinfo{person}{Gabriel Goh}, \bibinfo{person}{Sandhini Agarwal}, \bibinfo{person}{Girish Sastry}, \bibinfo{person}{Amanda Askell}, \bibinfo{person}{Pamela Mishkin}, \bibinfo{person}{Jack Clark}, {et~al\mbox{.}}} \bibinfo{year}{2021}\natexlab{}.
\newblock \showarticletitle{Learning transferable visual models from natural language supervision}. In \bibinfo{booktitle}{\emph{International conference on machine learning}}. PmLR, \bibinfo{pages}{8748--8763}.
\newblock


\bibitem[Radford et~al\mbox{.}(2019)]%
        {Radford2019LanguageMA}
\bibfield{author}{\bibinfo{person}{Alec Radford}, \bibinfo{person}{Jeff Wu}, \bibinfo{person}{Rewon Child}, \bibinfo{person}{David Luan}, \bibinfo{person}{Dario Amodei}, {and} \bibinfo{person}{Ilya Sutskever}.} \bibinfo{year}{2019}\natexlab{}.
\newblock \showarticletitle{Language Models are Unsupervised Multitask Learners}.
\newblock
\urldef\tempurl%
\url{https://api.semanticscholar.org/CorpusID:160025533}
\showURL{%
\tempurl}


\bibitem[Ren et~al\mbox{.}(2024)]%
        {RLMRec}
\bibfield{author}{\bibinfo{person}{Xubin Ren}, \bibinfo{person}{Wei Wei}, \bibinfo{person}{Lianghao Xia}, \bibinfo{person}{Lixin Su}, \bibinfo{person}{Suqi Cheng}, \bibinfo{person}{Junfeng Wang}, \bibinfo{person}{Dawei Yin}, {and} \bibinfo{person}{Chao Huang}.} \bibinfo{year}{2024}\natexlab{}.
\newblock \showarticletitle{Representation learning with large language models for recommendation}. In \bibinfo{booktitle}{\emph{Proceedings of the ACM Web Conference 2024}}. \bibinfo{pages}{3464--3475}.
\newblock


\bibitem[Rendle et~al\mbox{.}(2009)]%
        {BPR2009}
\bibfield{author}{\bibinfo{person}{Steffen Rendle}, \bibinfo{person}{Christoph Freudenthaler}, \bibinfo{person}{Zeno Gantner}, {and} \bibinfo{person}{Lars Schmidt-Thieme}.} \bibinfo{year}{2009}\natexlab{}.
\newblock \showarticletitle{BPR: Bayesian personalized ranking from implicit feedback}. In \bibinfo{booktitle}{\emph{Proceedings of the Twenty-Fifth Conference on Uncertainty in Artificial Intelligence}}. \bibinfo{pages}{452–461}.
\newblock


\bibitem[Rendle et~al\mbox{.}(2012)]%
        {MF}
\bibfield{author}{\bibinfo{person}{Steffen Rendle}, \bibinfo{person}{Christoph Freudenthaler}, \bibinfo{person}{Zeno Gantner}, {and} \bibinfo{person}{Lars Schmidt-Thieme}.} \bibinfo{year}{2012}\natexlab{}.
\newblock \showarticletitle{BPR: Bayesian personalized ranking from implicit feedback}.
\newblock \bibinfo{journal}{\emph{arXiv preprint arXiv:1205.2618}} (\bibinfo{year}{2012}).
\newblock


\bibitem[Schuster and Lazar(2024)]%
        {schuster2024attention}
\bibfield{author}{\bibinfo{person}{Nick Schuster} {and} \bibinfo{person}{Seth Lazar}.} \bibinfo{year}{2024}\natexlab{}.
\newblock \showarticletitle{Attention, moral skill, and algorithmic recommendation}.
\newblock \bibinfo{journal}{\emph{Philosophical Studies}} (\bibinfo{year}{2024}), \bibinfo{pages}{1--26}.
\newblock


\bibitem[Sheng et~al\mbox{.}(2024)]%
        {sheng2024language}
\bibfield{author}{\bibinfo{person}{Leheng Sheng}, \bibinfo{person}{An Zhang}, \bibinfo{person}{Yi Zhang}, \bibinfo{person}{Yuxin Chen}, \bibinfo{person}{Xiang Wang}, {and} \bibinfo{person}{Tat-Seng Chua}.} \bibinfo{year}{2024}\natexlab{}.
\newblock \showarticletitle{Language Representations Can be What Recommenders Need: Findings and Potentials}.
\newblock \bibinfo{journal}{\emph{arXiv preprint arXiv:2407.05441}} (\bibinfo{year}{2024}).
\newblock


\bibitem[Sun et~al\mbox{.}(2019)]%
        {sun2019multi}
\bibfield{author}{\bibinfo{person}{Jianing Sun}, \bibinfo{person}{Yingxue Zhang}, \bibinfo{person}{Chen Ma}, \bibinfo{person}{Mark Coates}, \bibinfo{person}{Huifeng Guo}, \bibinfo{person}{Ruiming Tang}, {and} \bibinfo{person}{Xiuqiang He}.} \bibinfo{year}{2019}\natexlab{}.
\newblock \showarticletitle{Multi-graph convolution collaborative filtering}. In \bibinfo{booktitle}{\emph{2019 IEEE international conference on data mining (ICDM)}}. IEEE, \bibinfo{pages}{1306--1311}.
\newblock


\bibitem[Tennenholtz et~al\mbox{.}(2023)]%
        {tennenholtz2023demystifying}
\bibfield{author}{\bibinfo{person}{Guy Tennenholtz}, \bibinfo{person}{Yinlam Chow}, \bibinfo{person}{Chih-Wei Hsu}, \bibinfo{person}{Jihwan Jeong}, \bibinfo{person}{Lior Shani}, \bibinfo{person}{Azamat Tulepbergenov}, \bibinfo{person}{Deepak Ramachandran}, \bibinfo{person}{Martin Mladenov}, {and} \bibinfo{person}{Craig Boutilier}.} \bibinfo{year}{2023}\natexlab{}.
\newblock \showarticletitle{Demystifying embedding spaces using large language models}.
\newblock \bibinfo{journal}{\emph{arXiv preprint arXiv:2310.04475}} (\bibinfo{year}{2023}).
\newblock


\bibitem[van~den Oord et~al\mbox{.}(2013)]%
        {BoW_text_features_music_MF-2013}
\bibfield{author}{\bibinfo{person}{Aaron van~den Oord}, \bibinfo{person}{Sander Dieleman}, {and} \bibinfo{person}{Benjamin Schrauwen}.} \bibinfo{year}{2013}\natexlab{}.
\newblock \showarticletitle{Deep content-based music recommendation}. In \bibinfo{booktitle}{\emph{Advances in Neural Information Processing Systems}}.
\newblock


\bibitem[Wang et~al\mbox{.}(2019)]%
        {wang2019neural}
\bibfield{author}{\bibinfo{person}{Xiang Wang}, \bibinfo{person}{Xiangnan He}, \bibinfo{person}{Meng Wang}, \bibinfo{person}{Fuli Feng}, {and} \bibinfo{person}{Tat-Seng Chua}.} \bibinfo{year}{2019}\natexlab{}.
\newblock \showarticletitle{Neural graph collaborative filtering}. In \bibinfo{booktitle}{\emph{Proceedings of the 42nd international ACM SIGIR conference on Research and development in Information Retrieval}}. \bibinfo{pages}{165--174}.
\newblock


\bibitem[Wu et~al\mbox{.}(2021b)]%
        {Wu2021RethinkingIH}
\bibfield{author}{\bibinfo{person}{Chuhan Wu}, \bibinfo{person}{Fangzhao Wu}, {and} \bibinfo{person}{Yongfeng Huang}.} \bibinfo{year}{2021}\natexlab{b}.
\newblock \showarticletitle{Rethinking InfoNCE: How Many Negative Samples Do You Need?}
\newblock \bibinfo{journal}{\emph{ArXiv}}  \bibinfo{volume}{abs/2105.13003} (\bibinfo{year}{2021}).
\newblock
\urldef\tempurl%
\url{https://api.semanticscholar.org/CorpusID:235212247}
\showURL{%
\tempurl}


\bibitem[Wu et~al\mbox{.}(2021a)]%
        {SGL2021}
\bibfield{author}{\bibinfo{person}{Jiancan Wu}, \bibinfo{person}{Xiang Wang}, \bibinfo{person}{Fuli Feng}, \bibinfo{person}{Xiangnan He}, \bibinfo{person}{Liang Chen}, \bibinfo{person}{Jianxun Lian}, {and} \bibinfo{person}{Xing Xie}.} \bibinfo{year}{2021}\natexlab{a}.
\newblock \showarticletitle{Self-supervised Graph Learning for Recommendation}. In \bibinfo{booktitle}{\emph{Proceedings of the 44th International ACM SIGIR Conference on Research and Development in Information Retrieval}}. \bibinfo{pages}{726–735}.
\newblock


\bibitem[Xi et~al\mbox{.}(2024)]%
        {KAR}
\bibfield{author}{\bibinfo{person}{Yunjia Xi}, \bibinfo{person}{Weiwen Liu}, \bibinfo{person}{Jianghao Lin}, \bibinfo{person}{Xiaoling Cai}, \bibinfo{person}{Hong Zhu}, \bibinfo{person}{Jieming Zhu}, \bibinfo{person}{Bo Chen}, \bibinfo{person}{Ruiming Tang}, \bibinfo{person}{Weinan Zhang}, {and} \bibinfo{person}{Yong Yu}.} \bibinfo{year}{2024}\natexlab{}.
\newblock \showarticletitle{Towards open-world recommendation with knowledge augmentation from large language models}. In \bibinfo{booktitle}{\emph{Proceedings of the 18th ACM Conference on Recommender Systems}}. \bibinfo{pages}{12--22}.
\newblock


\bibitem[Xu et~al\mbox{.}(2020)]%
        {xu2020reluplex}
\bibfield{author}{\bibinfo{person}{Jin Xu}, \bibinfo{person}{Zishan Li}, \bibinfo{person}{Bowen Du}, \bibinfo{person}{Miaomiao Zhang}, {and} \bibinfo{person}{Jing Liu}.} \bibinfo{year}{2020}\natexlab{}.
\newblock \showarticletitle{Reluplex made more practical: Leaky ReLU}. In \bibinfo{booktitle}{\emph{2020 IEEE Symposium on Computers and communications (ISCC)}}. IEEE, \bibinfo{pages}{1--7}.
\newblock


\bibitem[Yu et~al\mbox{.}(2023)]%
        {xsimGCL2023l}
\bibfield{author}{\bibinfo{person}{Junliang Yu}, \bibinfo{person}{Xin Xia}, \bibinfo{person}{Tong Chen}, \bibinfo{person}{Lizhen Cui}, \bibinfo{person}{Nguyen Quoc~Viet Hung}, {and} \bibinfo{person}{Hongzhi Yin}.} \bibinfo{year}{2023}\natexlab{}.
\newblock \showarticletitle{XSimGCL: Towards extremely simple graph contrastive learning for recommendation}.
\newblock \bibinfo{journal}{\emph{IEEE Transactions on Knowledge and Data Engineering}} (\bibinfo{year}{2023}), \bibinfo{pages}{913--926}.
\newblock


\bibitem[Zhang et~al\mbox{.}(2022)]%
        {bias-awareInfoNCE2022}
\bibfield{author}{\bibinfo{person}{An Zhang}, \bibinfo{person}{Wenchang Ma}, \bibinfo{person}{Xiang Wang}, {and} \bibinfo{person}{Tat-Seng Chua}.} \bibinfo{year}{2022}\natexlab{}.
\newblock \showarticletitle{Incorporating Bias-aware Margins into Contrastive Loss for Collaborative Filtering}. In \bibinfo{booktitle}{\emph{Advances in Neural Information Processing Systems}}. \bibinfo{pages}{7866--7878}.
\newblock


\bibitem[Zhang et~al\mbox{.}(2016)]%
        {knowledgebasedMF-2016}
\bibfield{author}{\bibinfo{person}{Fuzheng Zhang}, \bibinfo{person}{Nicholas~Jing Yuan}, \bibinfo{person}{Defu Lian}, \bibinfo{person}{Xing Xie}, {and} \bibinfo{person}{Wei-Ying Ma}.} \bibinfo{year}{2016}\natexlab{}.
\newblock \showarticletitle{Collaborative Knowledge Base Embedding for Recommender Systems}. In \bibinfo{booktitle}{\emph{Proceedings of the 22nd ACM SIGKDD International Conference on Knowledge Discovery and Data Mining}}. \bibinfo{pages}{353–362}.
\newblock


\bibitem[Zhang et~al\mbox{.}(2025)]%
        {zhang2025collm}
\bibfield{author}{\bibinfo{person}{Yang Zhang}, \bibinfo{person}{Fuli Feng}, \bibinfo{person}{Jizhi Zhang}, \bibinfo{person}{Keqin Bao}, \bibinfo{person}{Qifan Wang}, {and} \bibinfo{person}{Xiangnan He}.} \bibinfo{year}{2025}\natexlab{}.
\newblock \showarticletitle{Collm: Integrating collaborative embeddings into large language models for recommendation}.
\newblock \bibinfo{journal}{\emph{IEEE Transactions on Knowledge and Data Engineering}} (\bibinfo{year}{2025}).
\newblock


\end{thebibliography}

\appendix

\section{Hyperparameter Tuning for TextGCN-MLP}
\label{app:hyperparam}

\subsection{Algorithm}

In order to optimize the performance of the TextGCN-MLP model while keeping the computational cost manageable, we developed a two-stage hyperparameter tuning strategy. The process was decomposed into two distinct parts.

\subsubsection{Broad Partial Greedy Search}
\label{sec:broad_partial_search}

Since a full search over all hyperparameters would be time-consuming, we focused on tuning the most influential ones: the number of positive samples (\textit{n\_pos\_samples}), the number of negative samples (\textit{neg\_sample}), the learning rate (\textit{lr}), the output size of the MLP (\textit{output\_size}), and the number of graph convolution layers (\textit{n\_layers}).

In this stage, we fixed the default values (see Table~\ref{tab:final_params}). 
For each parameter, individual grid searches were performed using the ranges specified in Table~\ref{tab:broad_search}. 

\begin{table}[htbp]
\centering
\caption{Parameter ranges for the broad partial greedy search.}
\label{tab:broad_search}
\begin{tabular}{l|l}
\toprule
\textbf{Parameter} & \textbf{Values} \\
\midrule
\textit{output\_size} & \{16, 32, 64, 128, 256, 512, 1024\} \\
\textit{lr} & \{1e-5, 5e-5, 1e-4, 5e-4, 1e-3, 5e-3, 1e-2\} \\
\textit{n\_layers} & \{1, 2, 3, 4, 5, 8\} \\
\textit{neg\_sample} & \{16, 32, 64, 128, 256, 512, 1024\} \\
\bottomrule
\end{tabular}
\end{table}

\subsubsection{Narrow Full Grid Search}
\label{sec:narrow_full_search}

Based on the results of the broad search, we selected a subset of promising hyperparameters to perform a full grid search over all possible combinations. The search grid is detailed in Table~\ref{tab:grid_search}. All other parameters were kept at their default settings.

\begin{table}[t]
\centering
\caption{Hyperparameter grid for the narrow full grid search.}
\label{tab:grid_search}
\begin{tabular}{l|lll}
\toprule
\textbf{Parameter} & \textbf{Books} & \textbf{Video Games} & \textbf{Movies \& TV}\\
\midrule
\textit{n\_layers} & \{2, 3\} & \{2, 3, 4, 5\} & \{2, 3, 4, 5\}\\
\textit{output\_size} & \{128, 256\} & \{128, 256\} & \{128, 256\}\\
\textit{neg\_sample} & 512 & 512 & \{512, 2048, 4096\}  \\
\bottomrule
\end{tabular}
\end{table}

The grid search results, evaluated based on recall on the validation set, are summarized in Table~\ref{tab:final_params}.


\begin{table}[t]
\centering
\caption{Optimal hyperparameters (grid search) versus default values.}
\label{tab:final_params}
\begin{tabular}{llcc}
\toprule
Dataset & Parameter & Optimal & Default \\
\midrule
\multirow{4}{*}\textbf{Video Games} & \textit{lr} & 5e-4 & 5e-4 \\
                           & \textit{output\_size} & 128 & 64 \\
                           & \textit{neg\_sample} & 512 & 256 \\
                           & \textit{n\_layers} & 3 & 2 \\
\midrule
\multirow{4}{*}\textbf{Movies \& TV} & \textit{lr} & 5e-4 & 5e-4 \\
                           & \textit{output\_size} & 128 & 64 \\
                           & \textit{neg\_sample} & 512 & 256 \\
                           & \textit{n\_layers} & 2 & 2 \\
\midrule
\multirow{4}{*}\textbf{Books} & \textit{lr} & 5e-4 & 5e-4 \\
                           & \textit{output\_size} & 256 & 64 \\
                           & \textit{neg\_sample} & 512 & 256 \\
                           & \textit{n\_layers} & 3 & 2 \\
\bottomrule
\end{tabular}
\end{table}

\subsection{Model Hyperparameters}

\subsubsection{Number of Graph Convolution Layers}
\begin{figure}[t]
  \centering
  \includegraphics[width=0.7\linewidth]{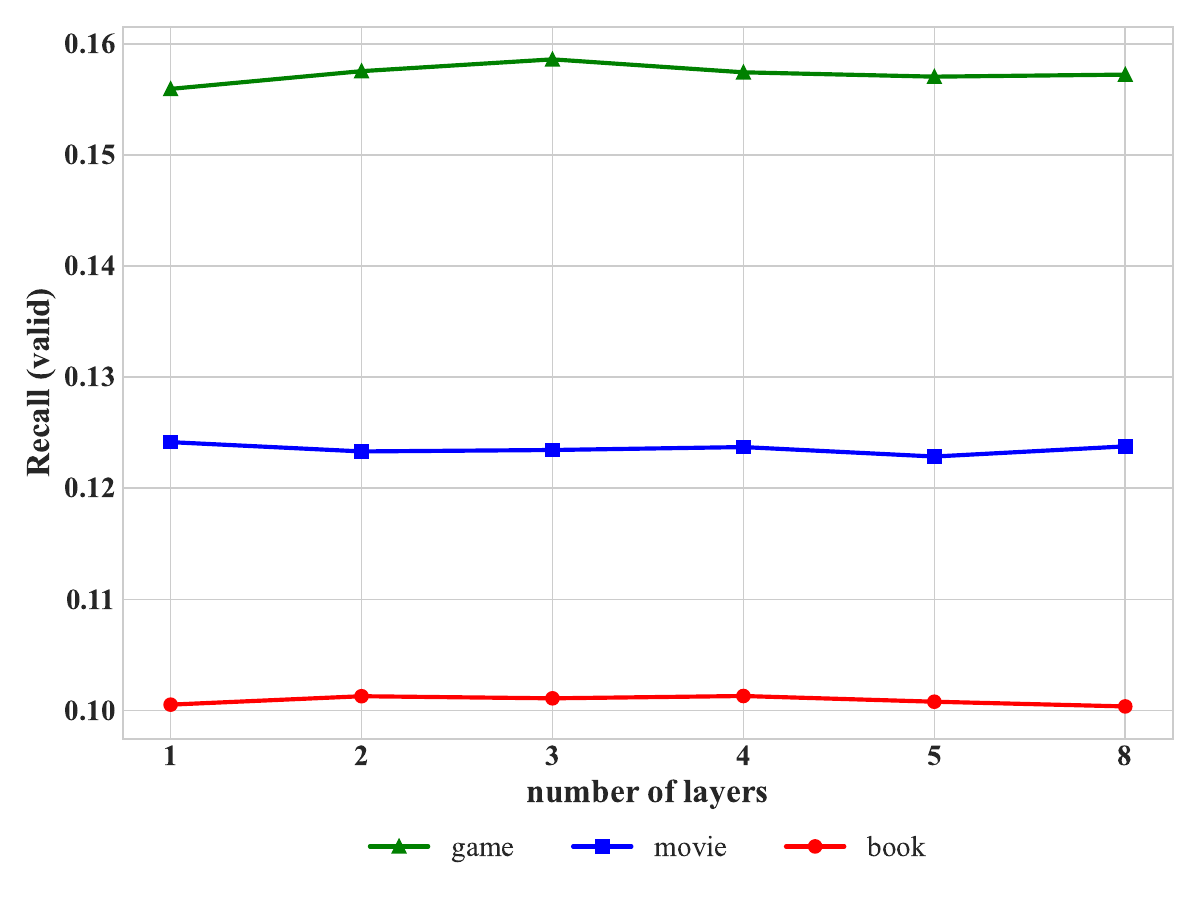}
  \caption{TextGCN-MLP partial dependence on the number of GCLs.}
  \label{fig:n_layers}
\end{figure}

The number of graph convolution layers has minimal impact on TextGCN-MLP performance (see Figure~\ref{fig:n_layers}). Specifically, using $2$, $3$, or $4$ layers yields comparable results.

\subsubsection{Output dimension of MLPs}
\begin{figure}[t]
  \centering
  \includegraphics[width=0.7\linewidth]{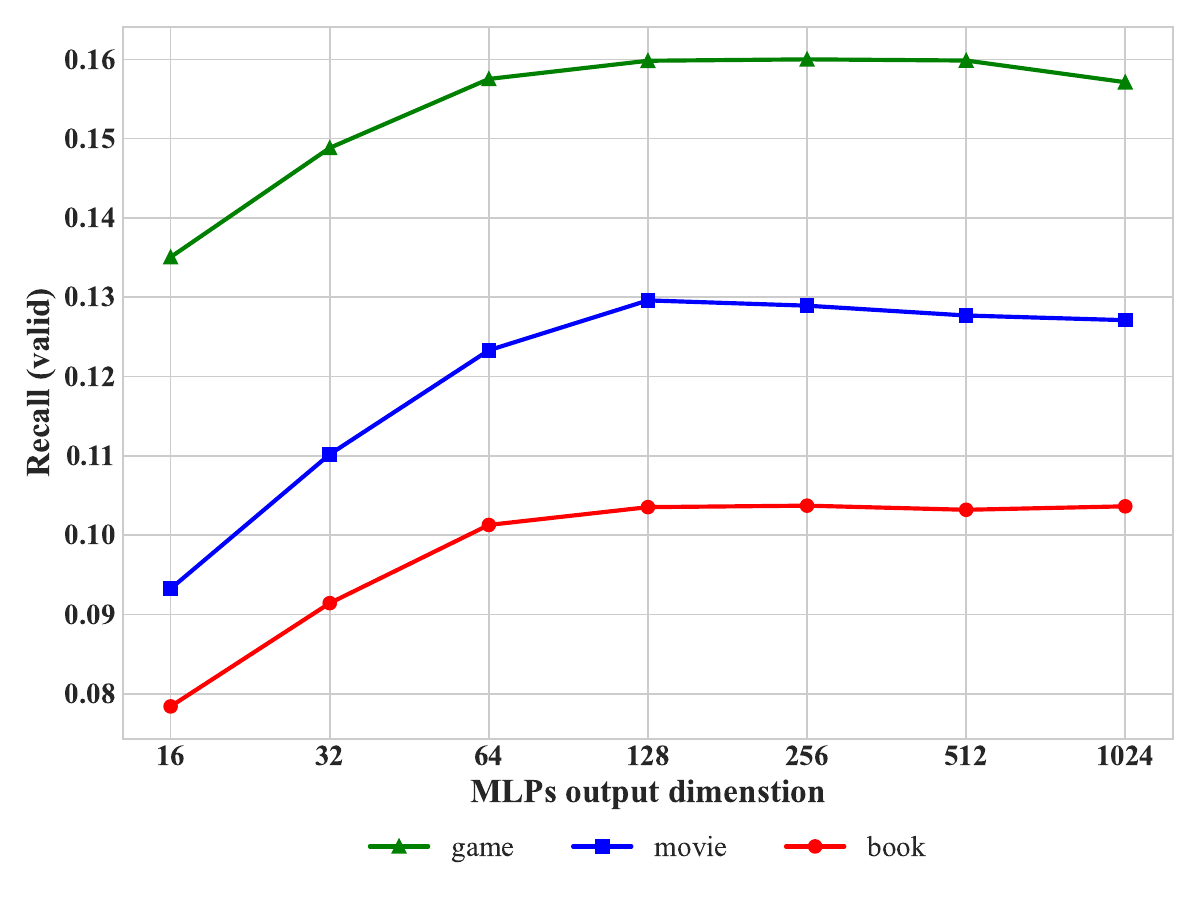}
  \caption{TextGCN-MLP partial dependence on the MLPs output size.}
  \label{fig:hidden_size}
\end{figure}

We find that the most impactful hyperparameter is the output dimension of the MLPs (see Figure~\ref{fig:hidden_size}). Increasing the output dimension to 128 provides a significant performance boost across all datasets compared to the default value of 64. Further increases beyond 128 do not lead to substantial improvements.

\subsection{Contrastive Loss Hyperparameters}

\subsubsection{Number of Negative Samples}
\begin{figure}[t]
  \centering
  \includegraphics[width=0.7\linewidth]{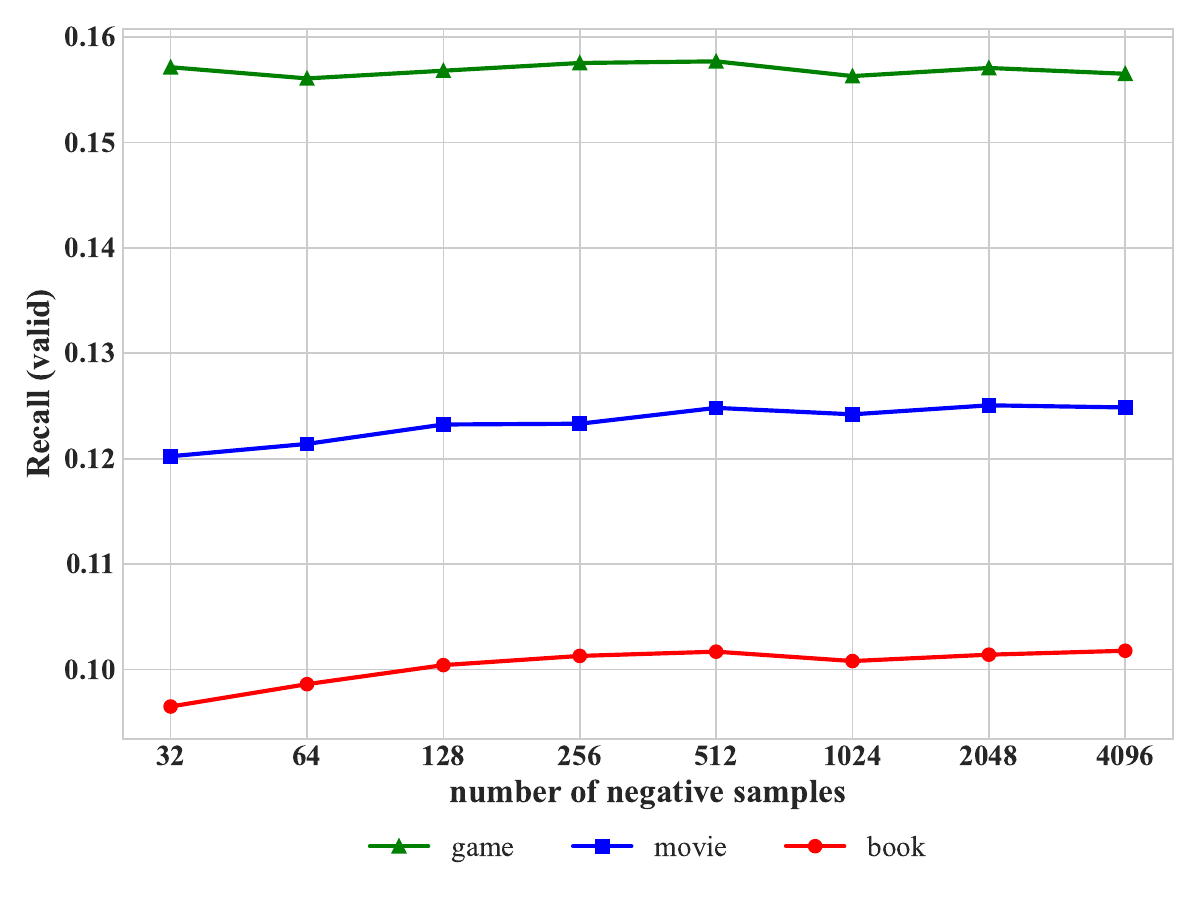}
  \caption{TextGCN-MLP partial dependence on the number of negative samples.}
  \label{fig:neg_sample}
\end{figure}

We observe a gradual improvement in performance as the number of negative samples increases. However, this gain plateaus or becomes negligible beyond 512 negatives (see Figure~\ref{fig:neg_sample}). We attribute this behavior to the inclusion of excessive noisy or false negatives when sampling beyond a certain threshold, a phenomenon also reported in prior studies on negative sampling for contrastive learning~\cite{Wu2021RethinkingIH}.

\subsubsection{Number of Positive Samples}
\begin{figure}[t]
  \centering
  \includegraphics[width=0.7\linewidth]{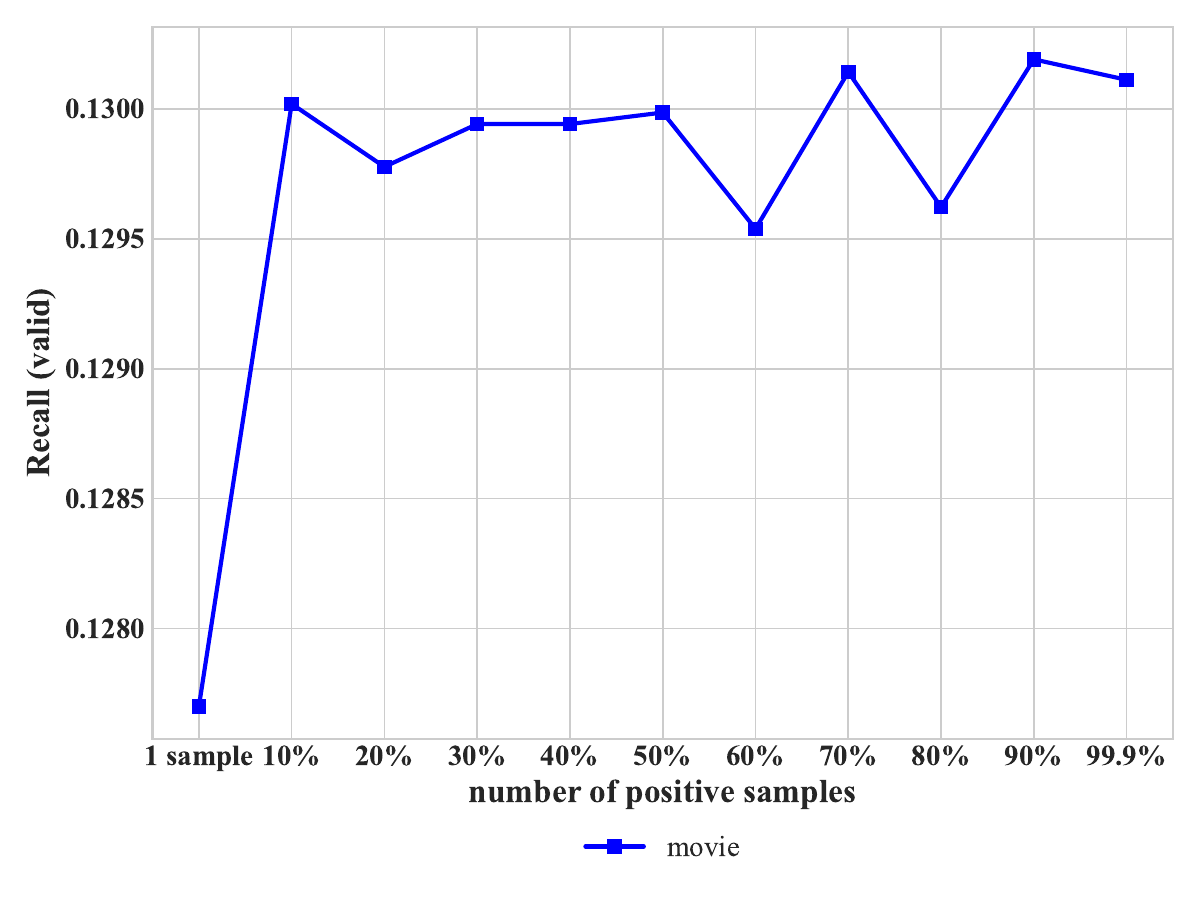}
  \caption{TextGCN-MLP partial dependence on the number of positive samples for the Movies dataset.}
  \label{fig:n_pos_samples_movie}
\end{figure}

Changing the number of positives from one to multiple consistently improved performance across datasets (see Figure~\ref{fig:n_pos_samples_movie}). This reaffirms our design choice of employing a k-positive contrastive loss in collaborative filtering. Furthermore, we observed varying degrees of improvement at different values of $k$, indicating that this hyperparameter should be tuned carefully. A closer examination of both validation and test sets across all datasets shows that the best gains tend to occur around the median percentile of user interactions. We attribute this to a balance between under-sampled and fully-sampled users near the median percentile range.

\begin{figure}[t]
  \centering
  \includegraphics[width=0.7\linewidth]{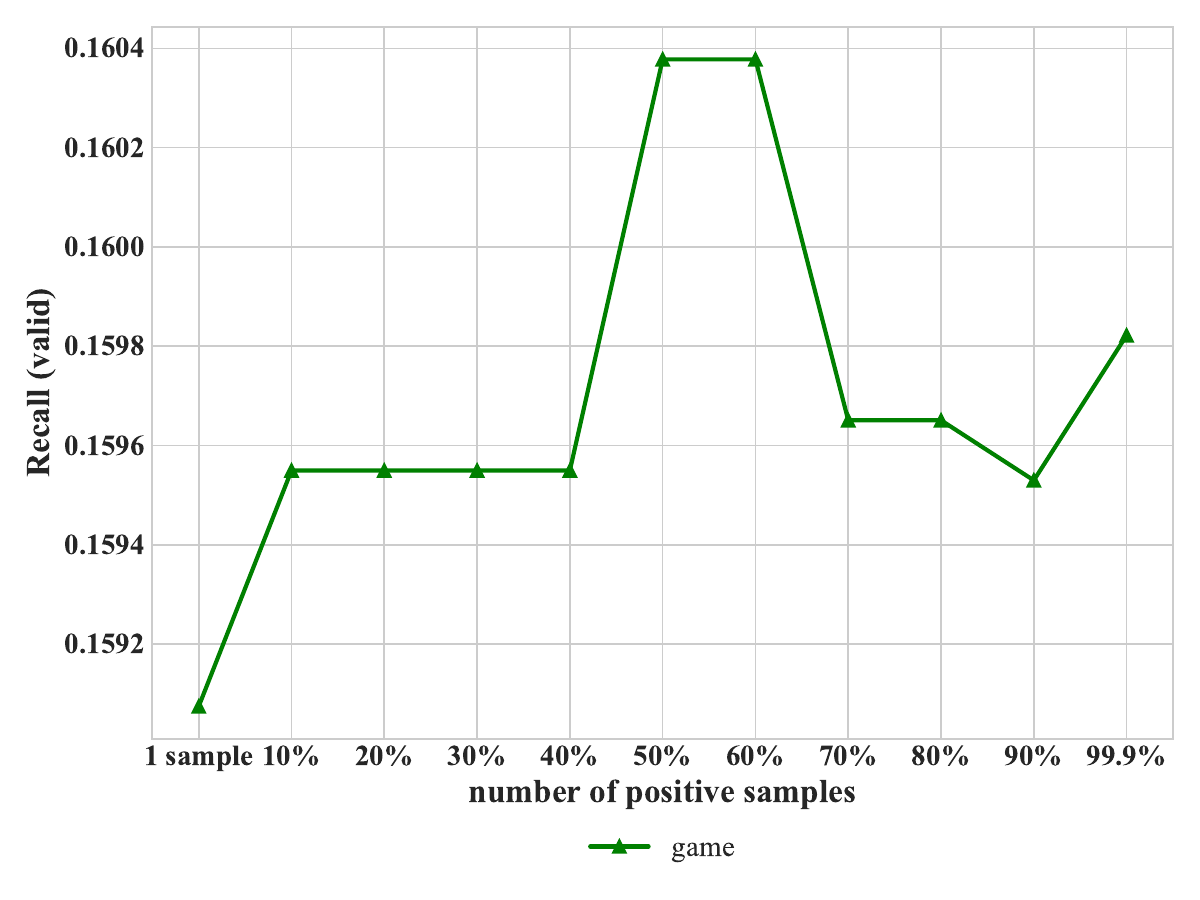}
  \caption{TextGCN-MLP partial dependence on the number of positive samples for the Video Games dataset.}
  \label{fig:n_pos_samples_game}
\end{figure}

\begin{figure}[t]
  \centering
  \includegraphics[width=0.7\linewidth]{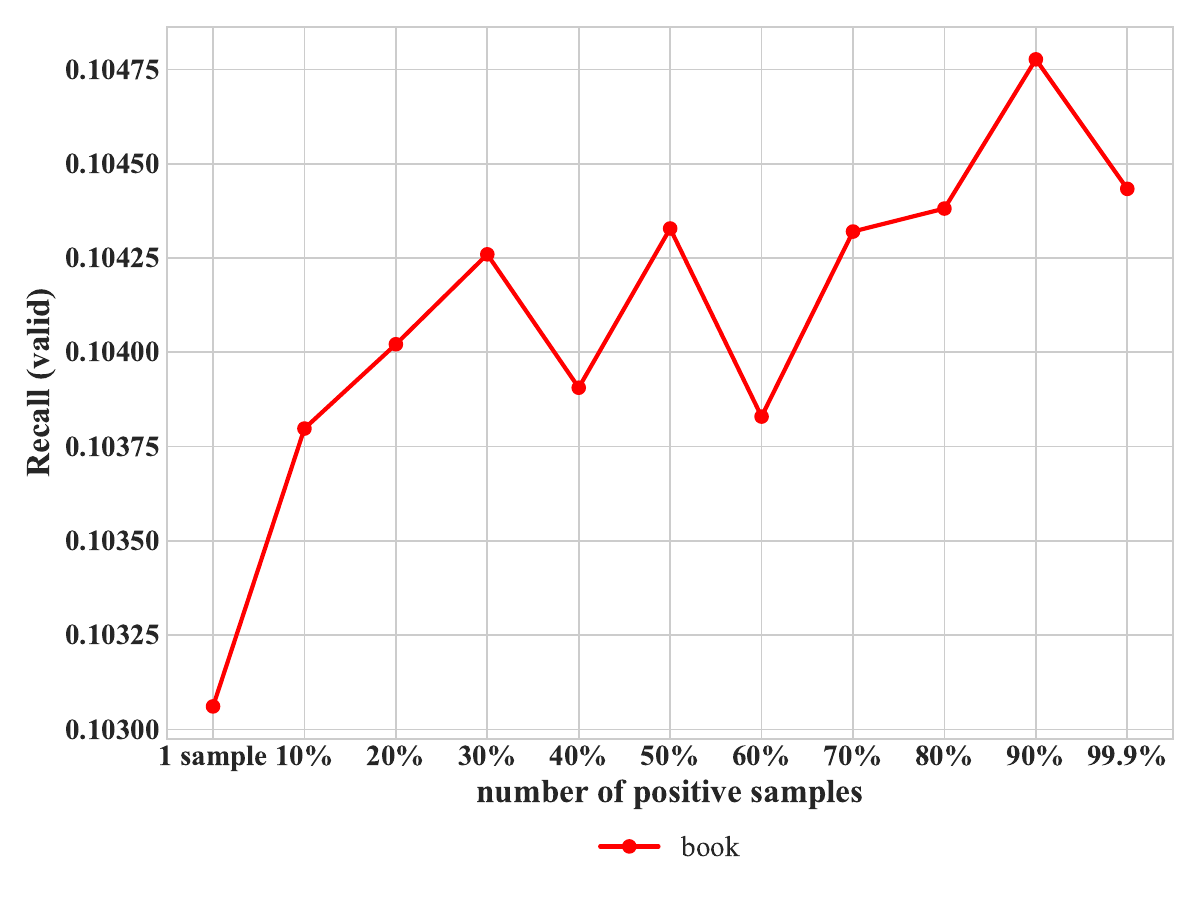}
  \caption{TextGCN-MLP partial dependence on the number of positive samples for the Books dataset.}
  \label{fig:n_pos_samples_book}
\end{figure}

\begin{figure}[t]
  \centering
  \includegraphics[width=0.7\linewidth]{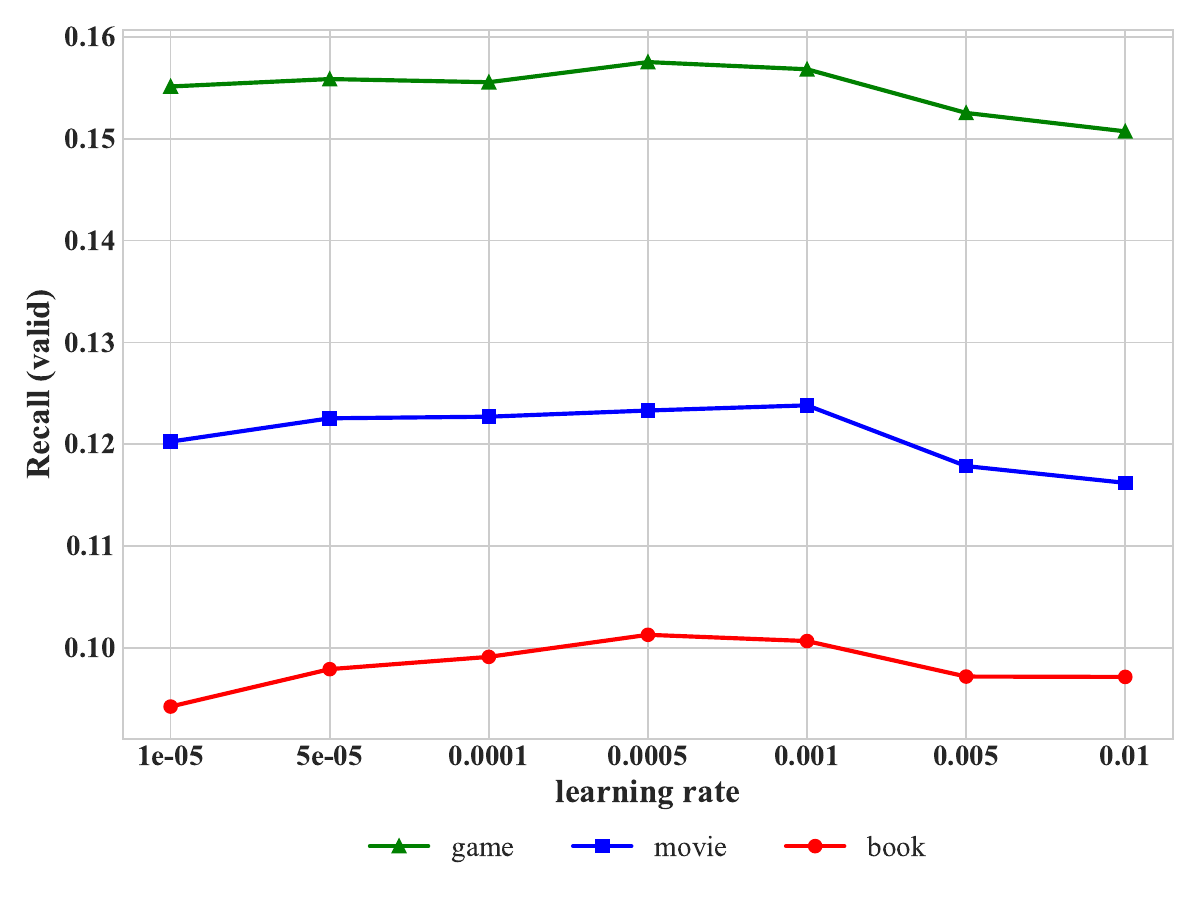}
  \caption{TextGCN-MLP partial dependence on the learning rate.}
  \label{fig:lr}
\end{figure}

\subsubsection{Learning Rate}
Regarding the learning rate, we observe two optimal values, $5\mathrm{e}{-4}$ and $1\mathrm{e}{-3}$, which yield approximately the same performance. Deviations from these values consistently decrease performance across all datasets (see Figure~\ref{fig:lr}.

\end{document}